\documentclass[a4paper,11pt]{article}
\pdfoutput=1

\usepackage{silence}  
\usepackage{jcappub} 
\usepackage[utf8]{inputenc}
\usepackage{amsfonts}
\usepackage{booktabs}
\usepackage{caption}
\usepackage{comment}
\usepackage{dcolumn} 
\usepackage{diagbox}
\usepackage{float}
\usepackage[all]{hypcap}
\usepackage{paralist}
\usepackage{physics}
\usepackage{siunitx}
\usepackage{slashed} 
\usepackage{subcaption}
\usepackage[dvipsnames]{xcolor}
\usepackage[normalem]{ulem}
\usepackage{cancel}

\DeclareSIUnit{\PeV}{\unit{\peta\eV}}
\DeclareSIUnit{\parsec}{\text{pc}}
\ExplSyntaxOn
\msg_redirect_name:nnn { siunitx } { physics-pkg } { none }
\ExplSyntaxOff
\usepackage[textsize=scriptsize,backgroundcolor=red!70,linecolor=red]{todonotes}

\newcommand{\setelems}[1]{\ensuremath{\qty{\numlist[list-final-separator={,\ }]{#1}}}}

\newcommand{\ie}{\textit{i.e.}}
\newcommand{\eg}{\textit{e.g.}}
\newcommand{\anti}[1]{\ensuremath{\bar{#1}}}
\newcommand{\zpr}{\ensuremath{Z^{\prime}}}
\newcommand{\zprn}[1]{\ensuremath{Z^{\prime({#1})}}}
\newcommand{\gpr}{\ensuremath{g^{\prime}}}
\newcommand{\Lmutau}{\ensuremath{L_\mu - L_\tau}}
\newcommand{\Umutau}{\ensuremath{U{(1)}_{\Lmutau}}}
\newcommand{\cnub}{\ensuremath{\text{C}\nu\text{B}}}
\newcommand{\gen}{IceCube-Gen\textit{2}}

\newcommand{\mkk}{\ensuremath{m_\text{KK}}}

\begin{document}

\title{Multiple-Mediator-Affected Ultra-High-Energy Neutrino Attenuation: Hints for 5D ${U(1)}_{L_\mu - L_\tau}$ in IceCube}

\author[a,b]{Atri Bhattacharya,}
\author[b]{Ayushi Kaushik,}
\author[b]{Kenji Nishiwaki}

\affiliation[a]{Space sciences,
  Technologies and Astrophysics Research (STAR) Institute,
  Université de Liège, Bât.~B5a, 4000 Liège,
  Belgium}
\affiliation[b]{Department of Physics, Shiv Nadar University, Tehsil Dadri,
  Gautam Buddha Nagar, Uttar Pradesh, 201314, India}

\emailAdd{A.Bhattacharya@uliege.be}
\emailAdd{ak356@snu.edu.in}
\emailAdd{kenji.nishiwaki@snu.edu.in}

\date{\today}
\abstract{
Recent IceCube observations point to the ultra-high energy neutrino flux
exhibiting a very soft spectral nature beyond tens of \unit{TeV}, which is
unexpected from the standard cosmic-ray neutrino connection.
As a potential explanation for this behaviour, we investigate neutrino
self-interactions in an extra-dimensional \( {U(1)}_{L_{\mu}-L_\tau} \) gauge
theory, where symmetry breaking leads to a tower of Kaluza-Klein gauge bosons in
compactified four dimensions.
These multiple gauge bosons may enhance the attenuation of astrophysical
neutrinos as they propagate in the C$\nu$B medium.
Unlike single-mediator scenarios, for example, which arise from broken
\( {U(1)}_{L_\mu - L_\tau} \) gauge symmetries in \emph{four} dimensions, the
presence of a Kaluza-Klein tower of mediators produces multiple closely spaced
resonances whose interference gives rise to a rich energy-dependent behaviour of
the scattering cross-section over a vast range of incident neutrino energies.
Alongside $s$-channel resonances, off-resonant $t$- and $u$-channel
contributions also become important.
We explore the possibility that repeated resonant scattering between
astrophysical neutrinos and those in the C$\nu$B, mediated by these new multiple
gauge bosons may increasingly attenuate the former at higher energies, thereby
addressing the possibility of softening its spectral nature within the
consideration of a single power law.
}

\maketitle

\section{Introduction}

Neutrinos are produced in astrophysical sources, which then propagate to Earth
traversing cosmological distances and resonantly scattering against cold, relic
neutrinos that form the cosmic neutrino background (\cnub).
The momenta transferred from the astrophysical neutrinos in such scattering
processes manifest as dips in their flux spectrum --- and consequently event
spectrum --- and should be detectable at Earth.
However, within the realm of the Standard Model~(SM), the mediator involved in
this scattering process is the \( Z \)-boson, which is far too heavy:
\( m_Z = \SI{91.2}{\GeV} \)~\cite{ParticleDataGroup:2026aaa}, and pushes the
resonance \( \qty(E^\text{res}_{\nu_i} = m^2_Z / 2m_i) \) (in the laboratory
  frame), where
\( m_i \text{ for } i \in \setelems{1;2;3} \)
are the neutrino masses, to well beyond \( \SI{100}{\PeV} \) and out of
statistical relevance for current large-volume neutrino telescopes.
Such a heavy mediator also makes the cross-section too weak to attenuate the
fluxes of these astrophysical, ultra-high energy (UHE) neutrinos via
interactions with the \cnub.
As a consequence, the flux of UHE neutrinos produced in astrophysical sources
remains unchanged in spectral shape as it traverses megaparsec distances to
reach Earth, propagating as if the medium were simply a vacuum.

The situation changes when considering physics beyond the Standard Model (BSM) that
accommodate light gauge bosons as part of their particle spectrum.
In particular, for BSM models with one or more \unit{\MeV} scale neutral gauge
bosons, the effect of resonant scattering on the astrophysical neutrino flux
spectrum may be seen at the ultra-high energies currently probed at
IceCube and expected, in the future, to be aided by future large volume
detectors like KM3NeT~\cite{KM3Net:2016zxf,KM3NeT:2025npi} and
\gen~\cite{IceCube-Gen2:2020qha}.

As an example, we consider BSM physics involving an anomaly-free broken \Umutau\ gauge symmetry,
where the spectrum of fundamental particles is supplemented by light neutral
gauge bosons \zpr\ which couple to the
{2\textsuperscript{nd} and 3\textsuperscript{rd} generation SM}
leptons~\cite{Foot:1990mn,He:1990pn,He:1991qd}.
In four dimensions, the breaking of \Umutau\ symmetry leads to the existence of
a single \zpr\ which may potentially have masses in the \unit{\MeV}
scale~\cite{Foot:1994vd,Asai:2017ryy,Asai:2018ocx}.
Models with such a broken symmetry display remarkably rich phenomenology,
especially in the neutrino sector~\cite{Asai:2017ryy,Asai:2018ocx} --- despite
being rather minimal extensions to the SM.
The recent discussions of the (minimal) \Umutau\ scenario are found, e.g., on
searching dark matter candidates via the interaction~\cite{Asai:2020qlp}, on the
profiles of active
neutrinos~\cite{Asai:2017ryy,Asai:2018ocx,Ibe:2025rwk,Ibe:2026yei}, to alleviate
tensions in cosmological
parameters~\cite{Escudero:2019gzq,Araki:2021xdk,Asai:2023ajh}, for their
implications for leptogenesis~\cite{Asai:2017ryy,Asai:2020qax}, and other
unexplained issues.\footnote{Such models were also used historically to
explain the muon magnetic moment anomaly~\cite{Baek:2001kca}, but with
recent theoretical computation of the value of \( g_\mu - 2 \) restoring
consistency with experimental observations~\cite{Boccaletti:2024guq,Aliberti:2025beg},
this will be less of a motivation going forward.
Refer to Section 1 of~\cite{Chakraborty:2024xxc} for constraints on the model
arising from other physical processes not discussed in this manuscript, as well
as for future predictions.}

On the other hand, in theories involving extra dimensions, the breaking of
\Umutau\ symmetry manifests in \( 4d \) as a tower of neutral gauge bosons with
proportionally increasing
masses~\cite{Chakraborty:2024xxc,Chakraborty:2025jbd,Chakraborty:2026ocj}, the
lightest mass being determined by the scale of the \emph{fifth} dimension.
In this scenario, astrophysical neutrinos may undergo resonant scattering at
multiple energies corresponding to the different mediator masses.
As a result, amplitudes corresponding to processes with the same initial and
final states, but different mediators, may constructively or destructively
interfere, manifesting as peaks and dips in the total cross-section.
These interference features, along with the presence of multiple resonance peaks
in the cross-section, distinguish the multiple mediator scenario
from the single light mediator case.
Furthermore, the consideration of off-resonant processes involving $t$-
and $u$-channels, which may wash out destructive interference dips in the
cross-section, become very important in the multiple mediator case when
computing the evolution of the flux from source to Earth.

The consequences of a single light gauge boson in four-dimensional ($4d$) BSM
theories acting as the mediator for astrophysical neutrinos scattering against \cnub\ have
been explored in detail in past works,
e.g.,~\cite{Araki:2014ona,Kamada:2015era,Araki:2015mya,DiFranzo:2015qea,Chauhan:2018dkd,Barenboim:2019tux,Bustamante:2020mep,Carpio:2021jhu,Hooper:2023fqn,Hyde:2023eph,Francener:2024bfm,delaVega:2024pbk,Francener:2025apz}
in the context of UHE neutrino events.
In this work, to explore the effect of interference terms in the same process
involving multiple mediators, we will study the case of \Umutau\ gauge symmetry
breaking in the five dimensional \( \qty(5d) \) Kaluza-Klein (KK) tower.
Corresponding to peaks and dips in the neutrino self-interaction
($\nu$SI) cross-section
mediated by a tower of \zprn{n}, in such a scenario, for appropriate mediator
masses, the UHE neutrino flux will experience a series of dips
and rises, respectively, with the relative height of the
latter dependent on how strongly $t$ and $u$ channel contributions wash them
away.
This imbues the UHE neutrino flux spectrum with very distinctive features and
makes such models testable at \SI{1}{\cubic\km} or larger volume neutrino
detectors looking for extragalactic, astrophysical neutrinos at energies
exceeding \SI{100}{\TeV}, including at the currently operational detector at the
South Pole --- IceCube~\cite{Halzen:2010yj}.

Operational since 2010, IceCube has collected more than a decade's worth of
statistically significant astrophysical neutrino events at energies upward of
\SI{10}{\TeV}~\cite{IceCube:2013low,IceCube:2013cdw,Abbasi:2021qfz,
  IceCube:2023sov,IceCube:2025tgp}.
When assuming the neutrino flux reaching Earth at these energies follows a
simple power-law spectrum, \( \phi_{\nu} \propto E_{\nu}^{-\gamma} \), these
events have consistently favoured explanations involving spectral indices
($\gamma$) characterising a steep fall-off with energy.
This is reflected in the relatively high best-fit value of the spectral index
\( \gamma = 2.84^{+0.11}_{-0.09} \) at \( E_\nu \geqslant \SI{60}{\TeV} \) for
the incident neutrino flux when considering the High Energy Starting Events
(HESE) sampling involving those events whose initial \( \nu N \) interaction
happens inside the detector instrumented
volume~\cite{IceCube:2020wum,IceCube:2023sov}.
Other event samplings at the same detector, based on differing event
morphologies or energy thresholds, suggest slightly different --- yet still
steeply falling --- best-fit incident
fluxes~\cite{Abbasi:2021qfz,IceCube:2024fxo,IceCube:2025tgp}.
For example, a fit to the starting muon track events with energies between
\SIrange{3}{550}{\TeV} finds \( \gamma = 2.58^{+0.10}_{-0.09} \)~\cite{IceCube:2024fxo}.
In addition, IceCube also sees a solitary Glashow resonance
event~\cite{IceCube:2021rpz} at energies of \SI{6.3}{\PeV}, with the initial
electronic interaction happening just outside the detector instrumented volume.
Even this solitary event provides a useful handle on the flavour
\( \qty(\phi_{\nu_e} : \phi_{\nu_\mu} : \phi_{\nu_\tau}) \) and
particle-antiparticle (\( \bar{\nu}_e: \nu_e \))~\cite{Skrzypek:2025tmg}
composition of the incident flux and, by extension, hints to the proton density
at/around the astrophysical body whence the neutrino flux
originates~\cite{Bhattacharya:2011qu,Muzio:2021zud,Bhattacharya:2023mmp}.

\vspace{1em}
UHE protons in the cosmic-ray flux are produced from the same
photo-hadronic interactions (see \eg,~\cite{Blandford:1987pw,Piran:2004ba} for
reviews) responsible for the production of UHE neutrinos~\cite{Halzen:2002pg}
at astrophysical sources.
Being electromagnetically charged, these protons repeatedly interact with,
amongst other intergalactic medium particles, photons constituting the cosmic
microwave background (CMB)~\cite{Berezinsky:2002nc}, and attenuate during their
journey to the Earth~\cite{Bhattacharjee:1999mup}.
The attenuated flux of protons makes up at least a fraction of the cosmic-ray
flux measured at Earth at \unit{\PeV} energies and higher, and fits a broken
power-law flux spectrum, with the spectral shape softening from
\numrange{2.7}{3.1} at the ``knee'' (see~\cite{Kachelriess:2019oqu} for a recent
review).
Assuming SM interactions of protons with intergalactic media as they
propagate to earth, this implies a starting flux with a much harder spectral
index \( \gamma \approx 2.0\text{--}2.2 \), which is also more in line with
expectations from Fermi shock acceleration~\cite{Fermi:1949ee} theories used to
predict high-energy particle production spectra at such sources.
It is, thus, natural to infer that UHE \emph{neutrinos}, produced in the
same interaction chain~\cite{Waxman:1997ti,Waxman:1998yy}, should have a
similarly hard spectral behaviour at the source.
However, within the purview of the SM, in contrast to protons,
neutrinos interact far too weakly with the intervening medium, \eg\ neutrinos
in the \cnub, and their fluxes do not attenuate as they propagate to Earth.
This implies that, in contrast to the cosmic-ray proton flux, the UHE neutrino
flux should arrive at Earth featuring a hard spectral shape similar to that at its production site.
The measurement of a much softer UHE neutrino flux spectrum at IceCube belies these
expectations.

BSM scenarios that allow enhanced neutrino self-interactions with increasing
energies may disproportionately attenuate astrophysical neutrino fluxes at
higher energies via scattering against neutrinos in the \cnub.
Consequently, despite originating with a relatively hard spectral index
\( \gamma \approx 2.0 \), UHE neutrino fluxes reaching Earth will exhibit
a softer spectral index, \( \gamma \approx 2.5\text{--}3.0 \).
This, then, may be a physics explanation for the softening
of the neutrino flux spectral shape at \unit{\PeV} and higher energies, in
confirmation with the observed IceCube neutrino event spectrum at energies
beyond \( \sim \SI{60}{\TeV} \), whilst still preserving the cosmic-ray-neutrino
connection at source.

\vspace{1em}
This work is organised as follows.
In Section~\ref{sec:model}, we discuss the specifics of the model under
consideration.
In Section~\ref{sec:nuflux}, we compute the modification to a simple power-law
astrophysical flux due to scattering against relic neutrinos by solving the
requisite Boltzmann equations.
In Section~\ref{sec:ic-events}, we use the modified flux to compute IceCube {starting} vertex events for energies above \SI{60}{\TeV} for a few example
choices of parameter values.
Statistical analyses of the BSM affected event rates to determine favoured
regions of the model parameter space vis-à-vis observed 12-year High
Energy Starting Events data reported in~\cite{IceCube:2023sov} are discussed in
Section~\ref{sec:statanal}.
Finally, we conclude with a discussion of the impact of our inferences in Section~\ref{sec:discon}.

\begin{table}[htb]
  \centering
  \begin{tabular}{ c S S }
    \toprule
    Mass / \(\qty(\SI{e-3}{\eV})\)
     & {Normal Hierarchy}
     & {Inverted Hierarchy}
    \\
    \midrule
    \( m_{\nu_1} \)
     & 29.8
     & 52.3
    \\
    \( m_{\nu_2} \)
     & 31.1
     & 53.0
    \\
    \( m_{\nu_3} \)
     & 59.1
     & 14.7
    \\
    \bottomrule
  \end{tabular}
  \caption{\label{tab:numass}Neutrino masses consistent with oscillation data:
    \( \Delta m^{2}_\text{sol} = \SI{7.537e-5}{\eV^2} \) and
    \( \Delta m^{2}_\text{atm} = \SI{2.5e-3}{\eV^2} \), and the maximum value of
    the sum total of neutrino masses allowed by constraints from
    cosmology: \( \sum_{i=1}^{3} m_{\nu_i} \leqslant \SI{0.12}{\eV} \).}
  \label{tab:neutrino-mass-choice}
\end{table}

\paragraph{Note on the masses of active neutrinos:}
For the rest of this work, we set neutrino masses to values determined by
current best-fit solar and atmospheric mass-squared differences from oscillation
studies~\cite{Esteban:2024eli} and the maximum sum total of neutrino masses
allowed by cosmological constraints, notably Planck data at \SI{95}{\%}
confidence level~\cite{Planck:2018vyg}:
\( \sum_{i=1}^{3} m_{\nu_i} \leqslant \SI{0.12}{\eV} \) (see
  Table~\ref{tab:neutrino-mass-choice} for details).
This allows us to work in a regime where all relic neutrinos are
non-relativistic and relativistic spectral broadening
features~\cite{Wang:2025qap, Machado:2025ltu} in scattering cross-section resonances do not apply.
In particular, $s$-channel peaks in neutrino self-interaction cross-sections
mediated by BSM gauge bosons will be sharply peaked.
Analysis of the case where the lightest neutrino has a rest energy less than the
\cnub\ thermal energy, and where such spectral-broadening features become
important, is left for future work.

\paragraph{Convention:}
We use natural units $c=\hbar=k_\text{B}=1$.

\section{Model for extra-dimensional \texorpdfstring{\Umutau}{L\_mu-L\_tau}
  gauge symmetry}%
\label{sec:model}

\subsection{In four dimensions}%
\label{sub:xsec4d}

In four dimensions, scattering between high-energy astrophysical neutrinos and
relic neutrinos of the \cnub\ has been extensively studied in the
literature~\cite{DiFranzo:2015qea}.
The interaction process,
\begin{equation}
  \overset{(-)}{\nu_i} + \overset{(-)}{\nu_j} \rightarrow \overset{(-)}{\nu_k} + \overset{(-)}{\nu_l},
\end{equation}
is mediated by a new gauge boson \zpr\ associated with the anomaly-free \Umutau\
gauge symmetry~\cite{Foot:1990mn, He:1990pn, He:1991qd, Foot:1994vd}.
The \zpr\ boson acquires a mass when the symmetry breaks.
Here, lower-case Latin indices \(i,j,k,l \in \setelems{1;2;3} \) label the active
neutrino mass eigenstates, and the overline denotes Dirac conjugation.
Unless explicitly subscripted, final-state neutrinos represent mass eigenstates
for which appropriate sums or averages over flavours have been performed.

The interaction Lagrangian (in the gauge eigenstates) extending the SM with the \Umutau\ gauge symmetry is
given by

\begin{equation}
  \mathcal{L}_{\text{int}}
  =
  g^\prime \zpr_{\rho}
  \left[
    (\bar{\mu}\gamma^{\rho}\mu)
    -
    (\bar{\tau}\gamma^{\rho}\tau)
    +
    (\bar{\nu}_{\mu}\gamma^{\rho} P_L \nu_{\mu})
    -
    (\bar{\nu}_{\tau}\gamma^{\rho} P_L \nu_{\tau})
    \right],
\end{equation}
where \gpr\ denotes the gauge coupling associated with the \Umutau\ symmetry,
\( \zpr_{\rho} \) is the corresponding gauge boson field with the Lorentz
vector index
(\( \rho \in \setelems{0;1;2;3} \)),
$\gamma^\rho$ are the four-dimensional gamma matrices, and
\( P_L = ( \mathbb{I}_4 - \gamma_5)/2 \) is the left–handed chiral
projection operator with the four-by-four identical matrix (\( \mathbb{I}_4 \))
and the chirality matrix ($\gamma_5$).
We neglect the kinetic mixing between the \Umutau\ and the hyper-charge field
strengths.

The corresponding four-dimensional neutrino-antineutrino scattering cross
section, \textit{when only including the s-channel}, is given by\footnote{We note that our calculation of the cross section differs from that presented in Ref.~\cite{Francener:2025apz, DiFranzo:2015qea} by a factor of \( 1/16 \).}

\begin{align}
  \label{eq:nu_nu_cross_section}
  \sigma_{4\mathrm{d}}^{s\text{-only}}(\nu_i \anti{\nu}_j \rightarrow \nu \anti{\nu})
  &=
  \sum_{k,l=1}^{3}
  Q_{ji}^{\prime} Q_{ji}^{\prime\ast}
  Q_{kl}^{\prime} Q_{kl}^{\prime\ast} \,
  \frac{1}{16} \frac{{\gpr}^{4}}{3\pi}
  \frac{
    s
  }{ {\left(s - m_{\zpr}^{2}\right)}^{2} + m_{\zpr}^{2}\Gamma_{\zpr}^{2} } \notag \\
  &=
  Q_{ji}^{\prime} Q_{ji}^{\prime\ast}
  \frac{1}{16} \frac{2{\gpr}^{4}}{3\pi}
  \frac{
    s
  }{ {\left(s - m_{\zpr}^{2}\right)}^{2} + m_{\zpr}^{2}\Gamma_{\zpr}^{2} },
\end{align}
where
\begin{equation}
s = 2 E_{\nu(\bar{\nu}),\text{astro}} \, m_{\nu(\bar{\nu}),\text{target}}
    \label{eq:s-at-lab}
\end{equation}
is the Mandelstam variable denoting
the squared centre-of-mass energy in the laboratory frame,  $m_{\nu(\bar{\nu}),\text{target}}$ is the target relic antineutrino/neutrino mass, and $E_{\nu(\bar{\nu}),\text{astro}}$ is the incident astrophysical neutrino/antineutrino energy.
The parameters \( m_{\zpr} \) and \( \Gamma_{\zpr} \) represent the mass and
decay width of the mediator, respectively.
For the \Umutau\ gauge symmetry, the effective charge factor appearing in the
cross section satisfies
\begin{align}
  Q          &:= \text{diag}\left( 0, +1, -1 \right),                   &
  Q^{\prime} &:= U^\dagger Q U,                                         & \notag \\
  Q'_{kl}
    &=
        U^{*}_{\mu k} U_{\mu l} - U^{*}_{\tau k} U_{\tau l},  & 
             \sum_{k,l=1}^{3} Q_{kl}^{\prime} Q_{kl}^{\prime\ast} &= 2, & \notag \\
  \sum_{k=1}^{3} Q_{kl}^{\prime} Q_{kl'}^{\prime\ast}
    &=
        U_{\mu l} U^\ast_{\mu l'} + U_{\tau l} U^\ast_{\tau l'}, &
  \sum_{l=1}^{3} Q_{kl}^{\prime} Q_{k'l}^{\prime\ast}
    &=
        U^{*}_{\mu k} U_{\mu k'} + U^{*}_{\tau k} U_{\tau k'}, &
\end{align}
where the formulae for the components of the Hermitian matrix $Q'$ are derived by the unitary condition of the Pontecorvo-Maki-Nakagawa-Sakata~(PMNS) neutrino mixing matrix $U$.
The quantities \( U_{\alpha i} \) are elements of the PMNS matrix, where Greek
subscripts (\( \alpha, \beta,\ldots \in \lbrace e,\ \mu,\ \tau \rbrace \)) label
the lepton flavour indices.

\begin{figure}[htb]
  \centering
  \scalebox{0.7}{
      \begin{subfigure}{0.4\textwidth}
        \raisebox{30pt}{
            \includegraphics[width=\textwidth]{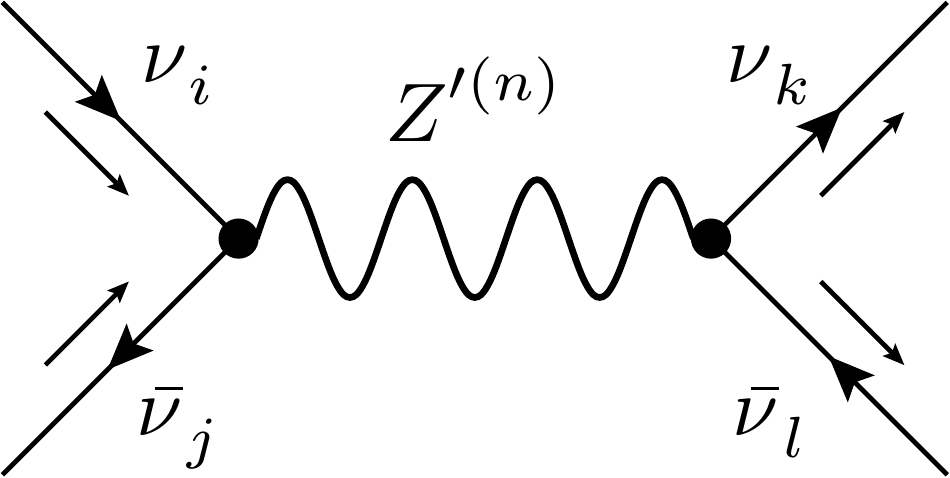}
        }
        \caption{}
      \end{subfigure}
      \hspace{15em}
      \begin{subfigure}{0.35\textwidth}
        \includegraphics[width=\textwidth]{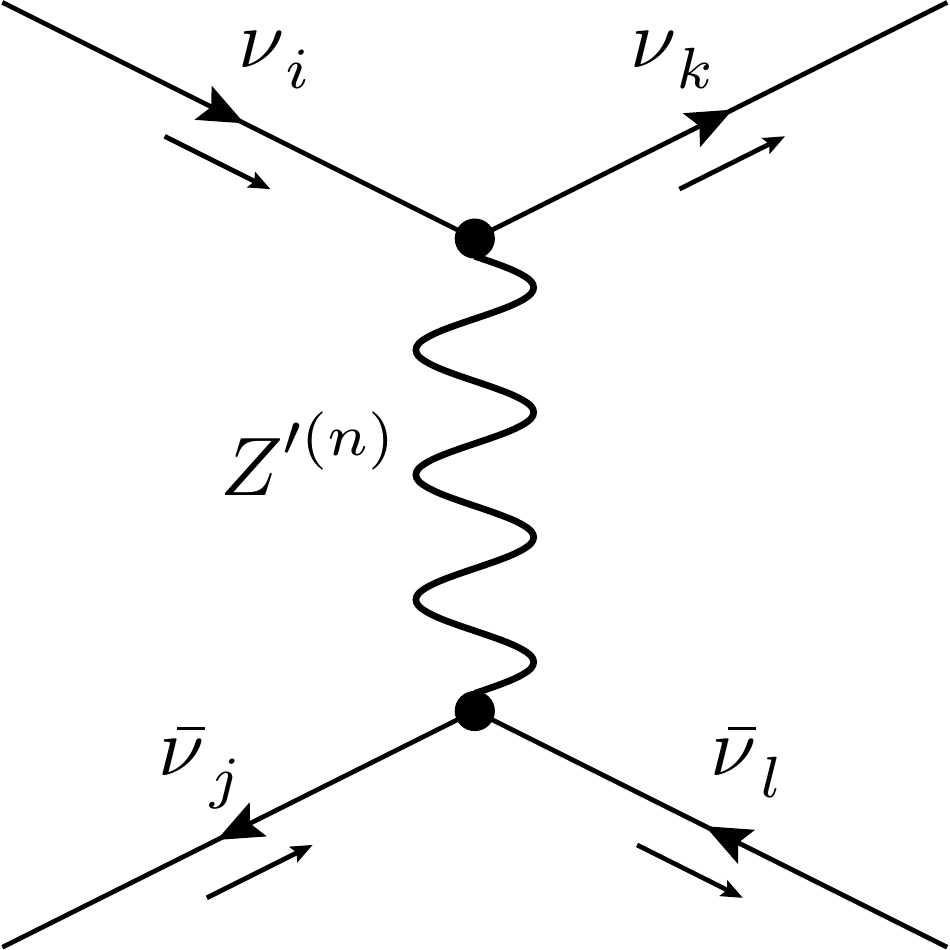}
        \caption{}
      \end{subfigure}
    }
    \\
    \vspace{20pt}
  \scalebox{0.7}{
      \begin{subfigure}{0.35\textwidth}
        \includegraphics[width=\textwidth]{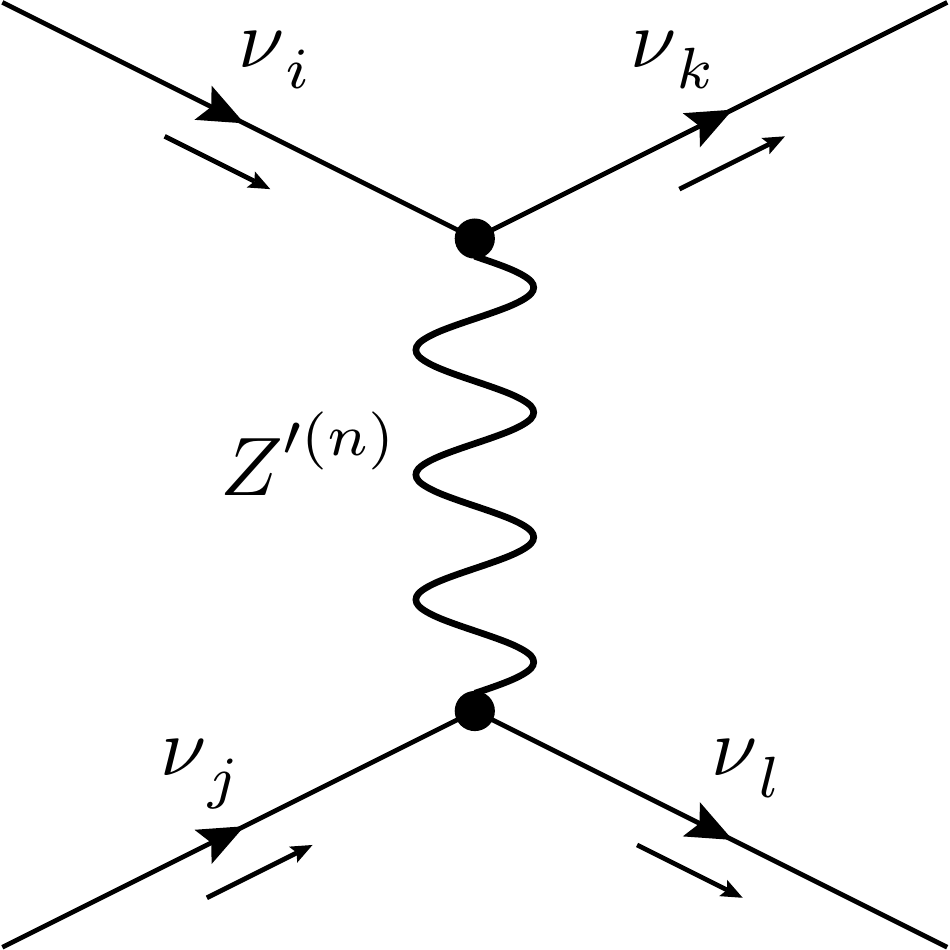}
        \caption{}
      \end{subfigure}
      \hspace{15em}
      \begin{subfigure}{0.35\textwidth}
        \includegraphics[width=\textwidth]{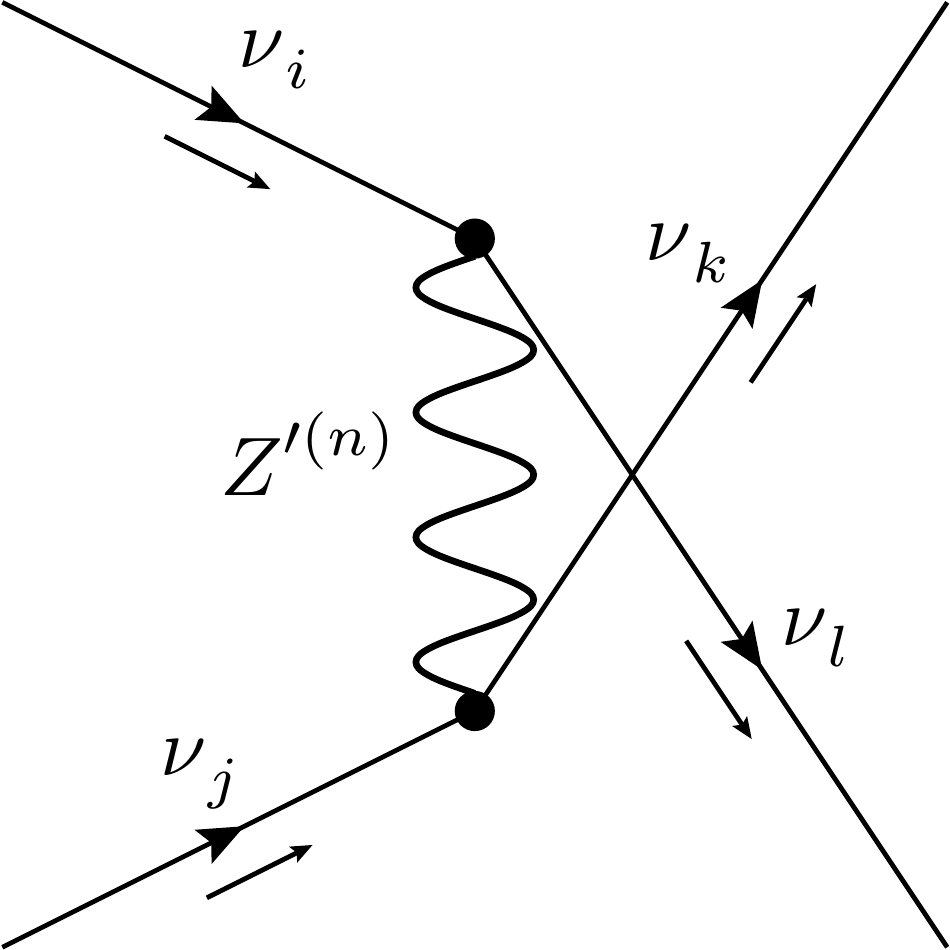}
        \caption{}
      \end{subfigure}
  }
  \caption{\label{fig:scatter-feyndiag}All possible tree-level diagrams
    contributing to \(\nu_i\) absorption (or, equivalently, to \(\nu_k\)
    regeneration), in 5 dimensions as mediated by the additional gauge
    bosons \zprn{n}:
    \(s\) and \(t\) channel diagrams for
    \(\nu \anti{\nu}\) scattering (top panel);
    \(t\) and \(u\) channel diagrams for \(\nu \nu\) scattering (bottom
    panel).
    We have ignored the contribution of diagrams with \zprn{n}\ boson
    final states.
  }
\end{figure}

If we would like to know the off-resonant structure precisely, we must also
include non-resonant channels:
\begin{inparaenum}[\itshape 1\upshape)]
  \item \(t\)-channel and its interference with the \(s\)-channel
  for \(\nu\bar{\nu}\) scattering, and
  \item both \(t\)- and \(u\)-channel scattering and their interference
  for \(\nu \nu\) scattering,
\end{inparaenum}
which are approximately as large as the \(s\) channel contribution away
from resonant energies.
All possible tree-level
diagrams contributing to the absorption of neutrino mass eigenstate \(\nu_i\)
is shown in Fig.\ \ref{fig:scatter-feyndiag}.

\subsection{Cross-sections in five dimensions}%
\label{sub:xsec5d}

We now extend the four-dimensional baseline established in Sec.~\ref{sub:xsec4d}
to a five-dimensional space-time framework, where the \Umutau\ gauge field
propagates in the extra dimension.
We eschew a full description of the resulting model and refer the
reader to 
Refs.~\cite{Chakraborty:2024xxc,Chakraborty:2025jbd,Chakraborty:2026ocj}
for details.
In a general $U(1)$ extension of this form,
kinetic mixing can naturally arise between the \Umutau\ gauge field and the
SM hyper-charge field. Throughout this work, however, we set the kinetic
mixing parameter to zero. This choice allows us to isolate the specific
phenomenological consequences of the extra-dimensional \Umutau\ interactions
without introducing additional, model-dependent electromagnetic couplings.

Within this framework, the scattering amplitudes for the extra-dimensional \Umutau\ processes shown in Fig.~\ref{fig:scatter-feyndiag} are given by:
\begin{align}
{\cal M}_{\nu_i \anti{\nu}_j \rightarrow \nu_k \anti{\nu}_l}
    &=
        {\sum_{n=1}^{N_\text{KKmax}}} \left[ \mathcal{M}_a^{(n)} + \mathcal{M}_b^{(n)} \right],\\
{\cal M}_{\nu_i \nu_j \rightarrow \nu_k \nu_l}
    &=
        {\sum_{n=1}^{N_\text{KKmax}}} \left[ \mathcal{M}_c^{(n)} + \mathcal{M}_d^{(n)} \right],
\end{align}
\begin{description}
  \item[s-channel for \(\nu\bar{\nu}\) case:]
        \begin{equation}
          \mathcal{M}_a^{(n)} = 
          - {Q'}_{ji} {Q'}_{k l}
          \frac{g_{\zprn{n}}^2 
          \left[ {\bar{v}_k} \gamma^{\mu} {{P}_{L}} {\bar{u}_l} \right]
          \left[ {\bar{u}_j} \gamma_{\mu} {{P}_{L}} {v_i} \right] }
          {s - m_{\zprn{n}}^2 + i\Gamma_{\zprn{n}} m_{\zprn{n}}}.
        \end{equation}
  \item[t-channel for \(\nu\bar{\nu}\) case:]
        \begin{equation}
          \mathcal{M}_b^{(n)} = 
          {Q'}_{ki} {Q'}_{j l}
          \frac{g_{\zprn{n}}^2 
          \left[ {\bar{v}_j} \gamma^{\mu} {{P}_{L}} {v_l} \right]
          \left[ {\bar{u}_k} \gamma_{\mu} {{P}_{L}} {u_i} \right] }
          { t - m_{\zprn{n}}^2 }.
        \end{equation}
  \item[t-channel for \(\nu\nu\) case:]
        \begin{equation}
          \mathcal{M}_c^{(n)} = 
          {{Q'}_{ki} {Q'}_{lj}}
          \frac{g_{\zprn{n}}^2 
          \left[ \bar{u}_l \gamma^{\mu} {{P}_{L}} u_j \right]
          \left[ \bar{u}_k \gamma_{\mu} {{P}_{L}} u_i \right] }
          { t - m_{\zprn{n}}^2 }.
        \end{equation}
  \item[u-channel for \(\nu\nu\) case:]
        \begin{equation}
          \mathcal{M}_d^{(n)} = 
          - {{Q'}_{li} {Q'}_{kj}} 
          \frac{g_{\zprn{n}}^2 
          \left[ \bar{u}_k \gamma^{\mu} {{P}_{L}} u_j \right]
          \left[ \bar{u}_l \gamma_{\mu} {{P}_{L}} u_i \right] }
          { u - m_{\zprn{n}}^2 },
        \end{equation}
\end{description}
where the categories (a) and (b) are for $\nu_i \anti{\nu}_j \rightarrow \nu_k
\anti{\nu}_l$, and (c) and (d) are for $\nu_i \nu_j \rightarrow \nu_k \nu_l$,
respectively.
\( u_i \) and \( v_i \) denote plane-wave solutions of $i$\textsuperscript{th}
Dirac neutrinos and antineutrinos where helicity indices are not explicitly stated.
$s$, $t$, and $u$ (without subscripts) represent Mandelstam variables.
Note that the relative minus signs between (a) and (b), also (c) and (d), originate from the Fermi statistics.
The \Umutau\ gauge interaction is characterised by the gauge coupling,
\begin{align}
 g_{\zprn{n}} &:= \gpr f^{(n)},&
 f^{(n)} &:= \cos(m_{\zprn{n}} y_{\text{SM}}),&
\end{align}
is the mode function and \(  y_{\text{SM}} \)
denotes the position of the SM brane in the bulk flat extra-dimensional space,
and \gpr\ is the effective 4d gauge coupling constant~\cite{Chakraborty:2024xxc}.
For all ensuing calculations, we use \( y_\text{SM} = 0 \) and
\( \gpr = \num{e-4} \).\footnote{
For the domain of $m_\text{KK}$ in
$\order{\numrange[range-phrase = \text{--}]{1}{10}}\,\unit{\MeV}$,
the current bound of NA64${}_\mu$ and the latest muon $(g-2)$ observation provide significant constraints on $g'$~\cite{Chakraborty:2025jbd}.
Our choice, $\gpr = \num{e-4}$, passes these bounds.
Note that another non-trivial constraint will come from $N_\text{eff}$, see~\cite{Escudero:2019gzq,Araki:2021xdk,Carpio:2021jhu,Asai:2023ajh} for examples in 4d; very roughly speaking, the case
$m_{\zpr} \lesssim \order{1}\,\unit{\MeV}$ (or a bit higher)  for
$\gpr \gtrsim  \num{5e-8}$ might be ruled out.
While similar constraints are expected for our five-dimensional model, there are
significant differences between it and conventional minimal four-dimensional
models---such as the presence or absence of an $SU(2)$ singlet scalar that
breaks the $\Umutau$ symmetry. Determining the precise constraints requires
analysing the corresponding coupled Boltzmann equations, a task that entails
extensive numerical analysis. Since the primary focus of this paper is the
behaviour of neutrino signals in scenarios involving multiple $Z'$ bosons, we
limit ourselves to a rough estimate regarding these constraints, leaving a
detailed evaluation for future research.
}

Furthermore, \( m_{\zprn{n}} \) and \( \Gamma_{\zprn{n}} \) correspond to the
mass and decay width of the \( n^{\text{th}} \) (\( n=1,2,3,\ldots \)) KK
excitation of the \zpr\ gauge boson, respectively.
The summation over $n$ accounts for contributions from all KK
modes propagating in the bulk.
Here, \( \mkk \) denotes the mass scale associated with the lowest KK
mode.
Additionally, the following definitions have been used:\footnote{
The form of the partial decay width, $\Gamma(Z^{'(n)} \to \nu_i \bar{\nu}_j) =
\frac{g_{\zprn{n}}^2}{ 24\pi} \abs{Q'_{ji}}^2 m_{\zprn{n}}$, is given
in~\cite{Chakraborty:2025jbd}.
In this case, if $Z'^{(n)}$ cannot kinematically decay into anything other than
neutrino pairs, the currently used expression follows.
Furthermore, in the case where $y_\text{SM} = 0$, the $n$-dependence drops out
of $g_{\zprn{n}}$, so the decay branching ratios between neutrinos become
independent of $n$.
}
\begin{subequations}
  \begin{align}
    \label{eq:ed_z_mass}
    m_{\zprn{n}} &= (2n - 1) \mkk\, \text{, and}
  \\
    \label{eq:ed_z_gamma}
    \Gamma_{\zprn{n}} &= \frac{
      g_{\zprn{n}}^2
    }{12 \pi} m_{\zprn{n}}\, \text{ for } n = 1, 2, 3\dots,
  \end{align}
\end{subequations}
where we assumed that $Z'^{(n)}$ cannot decay into anything other than neutrino pairs due to kinematics.
\(N_\text{KKmax}\) denotes a phenomenological cut-off for the KK summation since heavier modes tend to be decoupled more significantly.

We formulate the helicity-averaged differential cross sections analytically in the centre-of-mass frame first:
\begin{subequations}
  \begin{equation}
    \begin{aligned}
      \dv{\sigma_{\nu_i \bar{\nu}_j\rightarrow \nu_k \bar{\nu}_l}}{\Omega_\ast} =
      \frac{1}{64 \pi^2 s}
       & 
      \sum_{p=1}^{N_\text{KKmax}}
      \sum_{q=1}^{N_\text{KKmax}}
      g_{\zprn{p}}^2 g_{\zprn{q}}^2 \times
      \\
       &
      \left[\;
        \frac{u^2 |{Q'}_{ji}|^2 |{Q'}_{k l}|^2}{
          \left( s - m_{\zprn{p}}^2 - i\Gamma_{\zprn{p}} m_{\zprn{p}} \right)
          \left( s - m_{\zprn{q}}^2 + i\Gamma_{\zprn{q}} m_{\zprn{q}} \right)
        }
        \right.
      \\
       & \quad \left. + \;
        \frac{ u^2 |{Q'}_{ki}|^2 |{Q'}_{jl}|^2  }{
          \left( t - m_{\zprn{p}}^2 \right)
          \left( t - m_{\zprn{q}}^2 \right)}
        \right.
      \\
       & \quad
        \left. + \;
        \frac{ u^2 {Q'}_{ji} {Q'}_{kl} {Q'}_{ki}^* {Q'}_{jl}^* }{\left( t - m_{\zprn{p}}^2 \right)
          \left( s - m_{\zprn{q}}^2 + i\Gamma_{\zprn{q}} m_{\zprn{q}} \right)}
        \right.
      \\
       & \quad
        \left. + \;
        \frac{ u^2 {Q'}_{ji}^* {Q'}_{kl}^* {Q'}_{ki} {Q'}_{jl} }{\left( t - m_{\zprn{q}}^2 \right)
          \left( s - m_{\zprn{p}}^2 - i\Gamma_{\zprn{p}} m_{\zprn{p}} \right)}
        \;\right],
    \end{aligned}
  \end{equation}
  \text{for $\nu_i$-scattering against antineutrinos, and}
  \begin{equation}
    \begin{aligned}
      \dv{{\sigma}_{\nu_i \nu_j\rightarrow \nu_k \nu_l}}{\Omega_\ast}
      = \frac{1}{64 \pi^2 s}
       & 
      \sum_{p=1}^{N_\text{KKmax}}
      \sum_{q=1}^{N_\text{KKmax}}
      g_{\zprn{p}}^2 g_{\zprn{q}}^2 \times
      \\
       & \left[\;
        \frac{ {s^2 |{Q'}_{ki}|^2 |{Q'}_{lj}|^2} }{
          \left( t - m_{\zprn{p}}^2 \right)
          \left( t - m_{\zprn{q}}^2 \right)}
        +
        \frac{ {s^2 |{Q'}_{li}|^2 |{Q'}_{kj}|^2} }{
          \left( u - m_{\zprn{p}}^2 \right)
          \left( u - m_{\zprn{q}}^2 \right)}
        \right.
      \\
       & \quad \left.
        + \;
        \frac{ {s^2 {Q'}_{ki}^* {Q'}_{lj}^* {Q'}_{li} {Q'}_{kj}} }{
          \left( u - m_{\zprn{p}}^2 \right)
          \left( t - m_{\zprn{q}}^2 \right)}
        +
        \frac{ {s^2 {Q'}_{ki} {Q'}_{lj} {Q'}_{li}^* {Q'}_{kj}^*} }{
          \left( u - m_{\zprn{q}}^2 \right)
          \left( t - m_{\zprn{p}}^2 \right)}
        \; \right],
    \end{aligned}
    \label{eq:Cross_sec_ED}
  \end{equation}
  \text{for $\nu_i$-scattering against neutrinos.}
\end{subequations}
$\Omega_\ast$ denotes the solid angle in the centre-of-the-mass frame.
After the phase-space integration, we apply the Lorentz invariant $s$ variable in the lab frame as defined in~Eq.~\eqref{eq:s-at-lab}, so we get the total cross sections in the lab frame.
Note that $\mathrm{d}\sigma_{\nu_i \bar{\nu}_j\rightarrow \nu_k \bar{\nu}_l}/\mathrm{d}\Omega_\ast$ is straightforwardly obtainable from $\mathrm{d}\sigma_{\bar{\nu}_i \nu_j\rightarrow \bar{\nu}_k \nu_l}/\mathrm{d}\Omega_\ast$ by the exchanges, $i \leftrightarrow j$ and $k \leftrightarrow l$.
To get $\mathrm{d}\sigma_{\bar{\nu}_i \bar{\nu}_j\rightarrow \bar{\nu}_k \bar{\nu}_l}/\mathrm{d}\Omega_\ast$, one takes the complex conjugation for all of $Q'$ elements in $\mathrm{d}\sigma_{\nu_i \nu_j\rightarrow \nu_k \nu_l}/\mathrm{d}\Omega_\ast$.

In scenarios involving multiple mediators, such as a KK tower of 5d
gauge bosons, a complete description of the scattering process requires the inclusion of all relevant channels. While the $s$-channel contribution dominates
near resonances ($s \simeq m_{Z'^{(n)}}^2$), leading to Breit-Wigner
enhancements, the $t$- and $u$-channel diagrams provide non-resonant but
significant contributions, particularly at high energies.

In the presence of multiple KK modes, the $s$-channel amplitude involves a
coherent sum over propagators, which gives rise to interference effects. In
particular, destructive interference between different KK modes can produce
dip-like features in the total cross section.
However, when the $t$-channel contribution is included, it is smooth and
non-resonant behaviour partially compensates for these suppressions, effectively
filling in the dips generated by $s$-channel interference.
The $u$-channel is only viable for \( \nu \nu \) scattering, where the
$s$-channel scattering is not, and does not directly contribute to this wash-out
effect.

\begin{figure}[htb]
  \centering
  \includegraphics[width=0.85\linewidth]{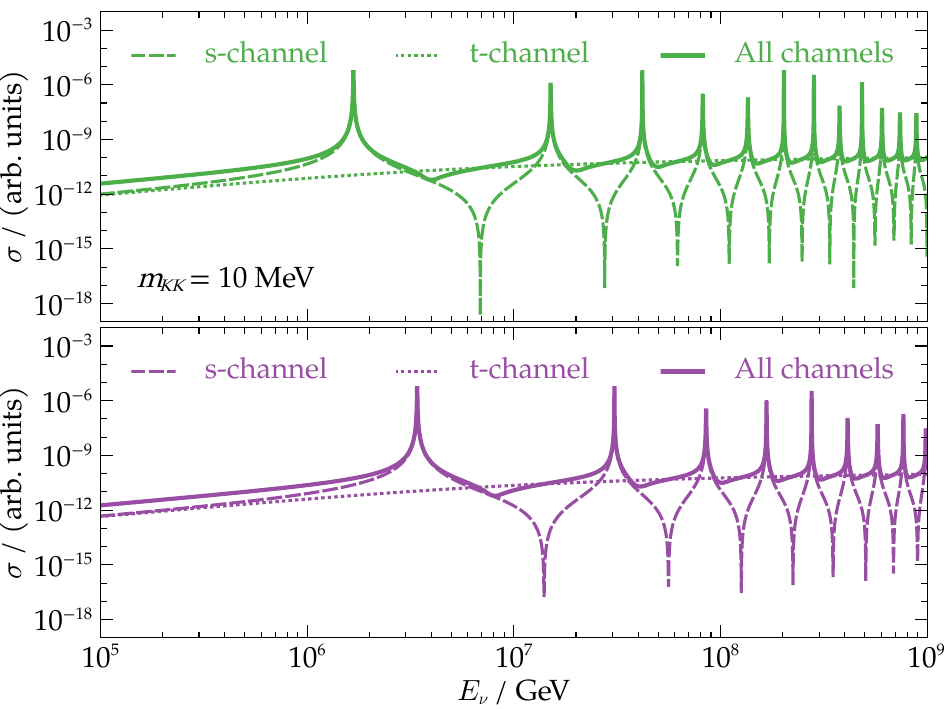}
  \caption{\label{fig:all-chans}Contributions to \( \nu\anti{\nu} \) scattering
    cross-section from $s$, $t$, and all channels combined --- $s$, $t$,
    \emph{and} $s$-$t$ interferences --- when using \( \mkk = \SI{10}{\MeV} \)
    for normal hierarchy (top) and inverted hierarchy (bottom).
    Final state spins have been summed over, and initial spins averaged to
    obtain a factor of \( 1/2 \).
    For each neutrino mass hierarchy, the lightest neutrino mass eigenstate is
    used as the \cnub\ target; \ie, \( m_{\nu_1} = \SI{29.8}{\meV} \)
    for NH and \( m_{\nu_3} = \SI{14.7}{\meV} \) for IH.
    The $s$-channel cross-section shows distinctive resonance peaks at
    \( \surd s = m_{\zprn{n}}\) as well as dips from the interference of
    $s$-channel amplitudes involving multiple \zprn{n}\ mediators.
    In the total cross-section, the $t$-channel cross-section without either of
    these effects largely wash out the interference dips from the $s$-channel
    interactions, leading to the distinctive nature demonstrated here.
    Note: We have set \( N_\text{KKmax} = 12 \), chosen to be large enough
      to demonstrate all resonant peaks up to the energy of \SI{e9}{\GeV} for
      the specific value of \mkk\ considered here.
  }
\end{figure}

In Fig.~\ref{fig:all-chans}, we present the total cross section for both normal
hierarchy (NH) and inverted hierarchy (IH), explicitly showing the individual
contributions from the $s$-channel, $t$-channel, and by taking all channels into
account. It is evident that the inclusion of the $t$-channel
significantly modifies the spectral structure, largely --- but not completely
--- washing out the dips from $s$-channel interference.

This highlights the importance of incorporating all channels in a consistent
manner. Although many 4$d$ studies of astrophysical neutrino self-interactions
focus primarily on the $s$-channel due to its resonant
enhancement~\cite{Francener:2025apz, DiFranzo:2015qea}, our results demonstrate
that in models with multiple mediators, such as the extra-dimensional framework
considered in this work, the $t$-channel and $u$-channel contributions, if viable, may not be neglected
and plays a crucial role in shaping the observable cross section and thus into
flux.

This behaviour has important implications for the astrophysical neutrino flux.
The overlapping of resonances leads to a continuous attenuation region, which
\emph{should} manifest as a broad absorption dip in the flux spectrum rather
than as discrete peaks. Consequently, this may result in a significant
suppression of detectable neutrino events above this energy range in experiments
such as IceCube.

\section{Astrophysical neutrino flux spectra}%
\label{sec:nuflux}

We model the starting neutrino flux summed over
all flavours and CP conjugate species as a single power-law:
\begin{equation}
  \label{eq:splflux}
  \phi^\text{ini}\qty(E_0) := \dv{N^\text{ini}}{E_0} = 
  \sum_{\alpha=e,\mu,\tau} \phi^\text{ini}_\alpha =
  \sum_{\alpha=e,\mu,\tau}
  \mqty(A_{\nu_\alpha} + A_{\bar{\nu}_\alpha}) \pqty{\frac{E_0}{\SI{100}{\TeV}} }^{-\gamma}\,,
\end{equation}
where
\( A := \sum_{\alpha=e,\mu,\tau} \mqty(A_{\nu_\alpha} + A_{\bar{\nu}_\alpha}) \) is the all-flavour normalisation constant pivoted at \SI{100}{\TeV} and
\( \gamma \) is the flavour-independent spectral index that quantifies the
steepness of the flux fall-off as a function of energy.
$A_{\nu_\alpha}$ and $A_{\bar{\nu}_\alpha}$ represent the magnitudes of initial flavour components, and we assume the initial neutrino and antineutrino flavour ratios to be
\( A_{\nu_e} : A_{\nu_\mu} : A_{\nu_\tau} = 1:1:0 \) and
\( A_{\bar{\nu}_e} : A_{\bar{\nu}_\mu} : A_{\bar{\nu}_\tau} = 0:1:0 \)
respectively, motivated by sources where neutrino production happens
dominantly from pion decays.
Note that $E_0$ characterises the energy of the initial flux of astrophysical neutrinos and antineutrinos, which should be connected to the redshifted energy at Earth,
\begin{equation}
    E_\nu := \frac{E_0}{1 + z},
    \label{eq:Enu-in-Earth}
\end{equation}
where $z$ is the redshift parameter.
Such a power-law assumption for the undisturbed neutrino flux is justified in
light of cosmic-ray observations at \SI{1}{\PeV} and higher energies, where the reconstructed spectrum follows either a simple or broken power law.
Current observations of the UHE neutrino spectrum at IceCube also fit well with
a simple power-law beyond energies of
\SI{60}{\TeV}~\cite{IceCube:2020wum,IceCube:2025tgp}, where the astrophysical neutrino flux starts to dominate over atmospheric neutrino background fluxes.

Notably, although a \( 4.7\sigma \) statistically significant preference for a
broken power law over a single power law is found for the medium energy starting
event (MESE) sampling used in~\cite{IceCube:2025tgp} between \SI{10}{\TeV} to
\SI{10}{\PeV}, we note that the best-fit value for the energy at which the
power-law changes its spectral shape is determined to be around \SI{30}{\TeV}.
When limiting any analysis exclusively to the high energy (HESE)
subset~\cite{IceCube:2020wum} --- events with deposited energies \( E_\text{dep}
\geq \SI{60}{\TeV} \) --- this broken power-law fit, therefore, is consistent
with a single-power law assumption favouring a very soft incident neutrino flux:
\( \gamma = 2.84^{+0.11}_{-0.09} \).

For our analysis, we allow \( \gamma \) for the starting neutrino flux to vary
between \numrange{2.0}{3.5}, with the best-fit being determined by fits to data,
in consonance with the flux normalisation \( A \).
As a consequence of the non-trivial nature of the scattering cross-section
involving astrophysical neutrinos interacting with the \cnub\ as discussed in
Section~\ref{sec:model}, the UHE neutrino flux eventually reaching the Earth
will have increasingly attenuated at higher energies and, therefore, be altered
in spectral shape and overall normalisation from its initial power-law nature.

Produced in their flavour eigenstates \( \ket{\nu_\alpha} \), neutrinos
propagate as mass eigenstates \( \ket{\nu_i} \), with the rotation between the
two bases described by the PMNS matrix:
\( \ket{\nu_\alpha} = \sum_{i} U_{\alpha i}^{\ast} \ket{\nu_i} \),
see e.g.,~\cite{Kayser:2004wmw}.
To understand the evolution of the spectral shape of the flux of neutrinos
(antineutrinos) in mass eigenstate \( \ket*{\nu_{i(\bar{i})}} \), in light of these BSM
interactions, we solve the Boltzmann
equations~\cite{Ng:2014pca,Araki:2015mya,Hooper:2023fqn,Francener:2025apz},
where the $z$-evolution of the fluxes of neutrino and antineutrino mass
eigenstates is calculated independently for each:
\begin{subequations}
  \label{eq:boltzmann}
\begin{equation}
  \begin{split}
    -(1+z)\,\frac{H(z)}{c}\,\dv{\phi_{i}}{z}
    &
    = J_{i}(E_0, z)
      - \phi_{i}(z) \sum_{j=1}^{3} \expval{n_{\nu_j}(z)\,\sigma_{ij}(E_0, z)}
    \\
    &
    \quad + \int_{E_0}^{\infty} \dd{E^{\prime}}
    \sum_{j,k=1}^{3} \phi_{j}(z) \expval{
      n_{\nu_k}(z) \,
      \sum_{l=1}^{3}P_{i\bar{l}}
      \dv{\sigma_{j\bar{k} \to i\bar{l}}(E^\prime, z)}{E_0}
    } \,,
  \end{split}
\end{equation}
\begin{equation}
  \begin{split}
    -(1+z)\,\frac{H(z)}{c}\,\dv{\phi_{\bar{i}}}{z}
    &
    = J_{\bar{i}}(E_0, z)
      - \phi_{\bar{i}}(z) \sum_{j=1}^{3} \expval{n_{\nu_j}(z)\,\sigma_{\bar{i}j}(E_0, z)}
    \\
    &
    \quad + \int_{E_0}^{\infty} \dd{E^{\prime}}
    \sum_{j,k=1}^{3} \phi_{j}(z) \expval{
      n_{\nu_k}(z) \,
      \sum_{l=1}^{3} {P_{l\bar{i}}}
      \dv{\sigma_{j\bar{k} \to {l\bar{i}}}(E^\prime, z)}{E_0}
    } \,,
  \end{split}
\end{equation}    
\end{subequations}
where \( \phi_{i(\bar{i})} \) represents the flux for the $i$\textsuperscript{th} neutrino (antineutrino) mass eigenstate.
Note that we impose the initial conditions for $\phi_{i(\bar{i})}$ as $\phi_{i(\bar{i})}(z=1)=0$, focusing on the situation that the objects originating the astrophysical neutrinos (and antineutrinos) are located at $z \simeq 1$.
The angle brackets mean the thermal average.\footnote{
$\phi_{i(\bar{i})}$ are functions of $z$, and $E_0$ (or $E_\nu$), though the dependence on the latter will not be explicitly written below. Furthermore, regarding the thermal average, we focus on the assumed situation in which $n_{\nu_i}$ are (nearly) constant, and the neutrino target mass is much larger than its momentum.
In this situation, as discussed in~\cite{Francener:2025apz}, the thermal average becomes trivial.
}

The redshift evolution of the astrophysical neutrino flux
\( \phi_{i(\bar{i})} \) is controlled by the term
\( -(1+z)\,H(z)c^{-1} \dd{\phi_{i(\bar{i})}}/\dd{z} \),
which describes the effect of cosmic expansion on propagation.
Here, $H(z)$ is the Hubble expansion rate as a function of the red shift $z$,
and for the red-shift range of interest, it can be approximated by
\begin{equation}
  \label{eq:hubble}
  H(z) \simeq H_0 \sqrt{\Omega_\Lambda + \Omega_M{\left(1+z\right)}^3}\,,
\end{equation}
where $H_0$ is the present value of the Hubble parameter,
$\Omega_M$ denotes the present-day matter density parameter, and
$\Omega_\Lambda$ represents the dark energy density parameter associated with
the cosmological constant, with $\Omega_\Lambda \simeq 1 - \Omega_M$; in our numerical analysis, we adopt the Planck best-fit values~\cite{Planck:2018vyg}:
$H_0 = \SI{67.4}{\km\per\s\per\mega\parsec}$ and $\Omega_M = 0.315$.

The first term on the R.H.S.\ of Eq.~\eqref{eq:boltzmann},
\( J_{i(\bar{i})}(E_0,z) \), denotes the astrophysical neutrino (antineutrino) 
production rate, which we parametrise as a single power-law
\begin{equation}
  \label{eq:iniflux}
  J_{i(\bar{i})}(E_0,z) := \phi_{i(\bar{i})}^\text{ini}(E_0) f(z)
  = A_{\nu_i(\bar{\nu}_i)} {\qty(\frac{E_0}{\SI{100}{\TeV}})}^{-\gamma} f(z)\,,
\end{equation}
where, in addition to quantities described in Eq.~\eqref{eq:splflux}, $f(z)$
describes the redshift distribution of the sources and
\( A_{\nu_i} = \sum_{\alpha=e, \mu, \tau} \abs{U_{\alpha i}}^2 A_{\nu_\alpha} \)
and
\( A_{\bar{\nu}_i} = \sum_{\alpha=e, \mu, \tau} \abs{U^\ast_{\alpha i}}^2 A_{\bar{\nu}_\alpha} \)
are the normalisations corresponding to the \(i\)\textsuperscript{th} neutrino
and antineutrino mass eigenstates, respectively.
We use the star formation rate (SFR) model~\cite{Hooper:2023fqn,Yuksel:2008cu} for \( f(z) \),
which peaks around \( z \simeq 1 \):
\begin{align}
    f(z) \propto \begin{cases}
      {(1+z)}^{3.4},                      \quad & z \leq 1 \\
      2^{3.7} \, {(1+z)}^{-0.3},          \quad & 1 < z < 4 \\
      2^{3.7} \,5^{3.2} \, {(1+z)}^{-3.5} \quad & z \geq 4, \\
            \end{cases}
\end{align}
where the normalisation is involved in $A_{\nu_i(\bar{\nu}_i)}$.

\begin{figure}[t]
  \centering
  \includegraphics[width=0.85\textwidth]{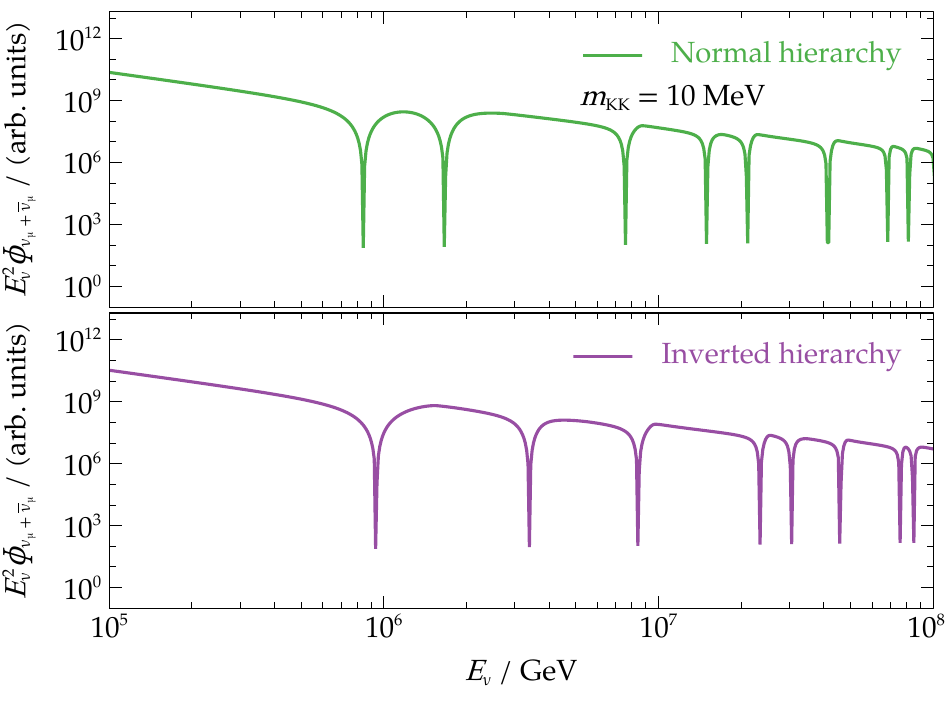}
  \caption{\label{fig:nuflux5d}
    Fluxes for the \( \nu_\mu + \bar{\nu}_{\mu} \) flavour at Earth, assuming
    \(\phi^\text{ini} \propto E^{-2.85}\)~[see Eq.~\eqref{eq:splflux}]
    injection with arbitrary normalisation (to be determined later by fitting to
    IceCube event rates) and allowing neutrino self-interactions while using
    normal (top) and inverted (bottom) hierarchies of neutrino masses.
    \(\mkk = \SI{10}{\MeV} \) has been used as a representative value of the
    mass of the lowest KK mode. Note that, for the injection flux, we have used
    the SM best-fit spectral index
    ($\gamma = 2.85$)~\cite{IceCube:2025cjy}, \ie\ one obtained without
    allowing for light gauge boson mediated neutrino self-interactions.
    We use \( N_\text{KKmax} = 10 \), big enough to demonstrate repeated
    resonance peaks up to energies of \SI{e8}{\GeV}.
  }
\end{figure}

The second term describes the depletion of the flux of neutrino species
\( \nu_i \) as it scatters against relic neutrinos \( \nu_j \) via the
interactions with
\( \sigma_{ij}
  = \sum_{k,l}
    (\sigma_{\nu_i \anti{\nu}_j \to \nu_k \anti{\nu}_l}
     + \sigma_{\nu_i \nu_j \to \nu_k \nu_l}) \),
where a sum is implied over all possible final neutrino mass eigenstates.
For the antineutrino species, we use
$\sigma_{\bar{i}j}
  = \sum_{k,l}
    (\sigma_{\anti{\nu}_i \nu_j \to \anti{\nu}_k \nu_l}
     + \sigma_{\anti{\nu}_i \anti{\nu}_j \to \anti{\nu}_k \anti{\nu}_l})
$.
We note that the neutrino-antineutrino scattering processes involve $s$- and
$t$-channel graphs, and therefore leads to peaked resonances, whereas the latter
neutrino-neutrino scattering processes only involve $t$- and $u$-channel graphs
with both exhibiting smooth behaviour.
The depletion rate is proportional to the number density of relic neutrinos in
the $j$ eigenstate \( n_{\nu_j} \) and the model specific self-interaction
cross-section.
For the neutrino number density in \cnub\ we use
\( n_{\nu_j} = n_{\bar{\nu}_j} = \SI{56}{\per\cubic\cm} \)~\cite{Schramm:1980xv}
for each neutrino and antineutrino species.\footnote{
Note that we used the condition $n_{\nu_j} = n_{\bar{\nu}_j}$ in Eq.~\eqref{eq:boltzmann}.
}

In a related vein, the third term describes interactions of UHE antineutrinos
\( \bar{\nu}_k \) against \cnub\ species \( \nu_j \) to regenerate \( \nu_i \) via the
following processes:
\( \sigma_{j\bar{k} \to i\bar{l}}
  = \sum_n \sigma(\nu_j \anti{\nu}_k \to \zprn{n} \to \nu_i \anti{\nu}_l)
\), and similarly for antineutrinos \( \bar{\nu}_i \)
as
$
\sigma_{j\bar{k} \to l\bar{i}}
= \sum_n \sigma(\nu_j \anti{\nu}_k \to \zprn{n} \to \nu_l \anti{\nu}_i)
$.
The branching fraction of \zprn{n}\ to specific neutrino mass eigenstates is
represented by \( P_{i\bar{l}} \), defined as an average of the branching ratios
$\{ \text{Br}\qty(\zprn{n} \to \nu_i \bar{\nu}_l) \}_{n=1,2,\ldots N_\text{KKmax}}$.
Although, the \( \zprn{n} \) may decay into charged lepton pairs
  \( \mu^{\pm} \) and \( \tau^{\pm} \), these processes only become viable for
  \( m_{\zprn{n}} \geqslant 2 m_\mu \).
  At these values, which represent higher $n$ KK modes, the resonance is pushed
  to energies beyond \( \order{100}\,\unit{\PeV} \), well out of the energy
  range considered for IceCube HESE analysis, which extends up to \SI{10}{\PeV}.
  Therefore, for all our calculations, we limit the maximum value of $n$ so that
  \( (2n - 1) \mkk < 2 m_\mu \); this implies for us, practically, the
  \(\zprn{n}\) can only decay to neutrino-antineutrino pairs and
  \( \sum_{i,l=1}^{3} P_{i\bar{l}} = 1 \).

The final two terms involve neutrino self-interaction cross-sections, and
involve BSM physics in the form of \zprn{n}\ mediated interactions that affect
changes in the shape of the flux \( \phi_i \) from the assumed injection
spectrum \( J_i(E_0, z) \).
We implement these cross sections in the
\texttt{nuSIprop}~\cite{Esteban:2021tub} code,
a Boltzmann equation solver tailored for neutrino self-interactions, to obtain
the propagated neutrino flux.

The final (anti)neutrino flux observed at Earth is found by integrating the \(
\dd{\phi_{i(\bar{i})}} / \dd{z} \) from Eq.~\eqref{eq:boltzmann} over the
redshift $z$, with the substitution for the redshifted energy at Earth,
$E_\nu$ as defined in Eq.~\eqref{eq:Enu-in-Earth}.

In Fig.~\ref{fig:nuflux5d}, we show the resulting flux for the muon neutrino
and antineutrino flavours combined, assuming an injection
spectrum \( \phi^\text{ini}_i(E_0) \propto E_0^{-2.85} \) with an arbitrary normalization
(to be fixed by fitting IceCube event rates).
In thus choosing the injection spectrum, we have set the spectral index to the
standard best-fit value obtained without considering the BSM-mediated neutrino
self-interactions~\cite{Ibe:2025rwk}.
We show the resulting fluxes for both the normal hierarchy (top panel) and the inverted hierarchy (bottom panel) of neutrino masses, for a representative value of \( \mkk = \SI{10}{\MeV} \).

The atypical nature of the resulting flux is a reflection of the peaks and dips
in the scattering cross-sections mediated by multiple \zprn{n}.
Corresponding to $s$-channel resonant peaks in the cross-sections, the flux
exhibits well-defined, narrow troughs.
Also, where there are dips in the former, due to interferences between multiple
mediator diagrams in the $s$-channel amplitudes, the flux shows discernible,
smooth peaks, which supplement the effects from regeneration in the Boltzmann
equations.
As noted previously (see Section~\ref{sec:model}), resonances at increasing
energies correspond to processes mediated by increasingly massive \zprn{n},
which, in turn, have increased decay widths [see Eq.~\eqref{eq:ed_z_gamma}].
Consequently, the troughs in the flux spectrum at high energies become
increasingly narrow and may start to overlap with each other.
In combination with the finite energy resolution of UHE neutrino detectors, the net consequence of this ``bunching up'' effect may be seen in terms of an overall
drop-off of event rates at increasingly higher energies rather than as well
delineated peaks and troughs.

The resulting flux of neutrino flavour \( \alpha \) at Earth after may then
be computed by rotating back from the mass eigenvectors.

\section{IceCube event rates}%
\label{sec:ic-events}

To compute HESE 12-year binned event predictions for the UHE neutrino flux arriving
at Earth after being scattered against \cnub\ via \zprn{n}\ mediators, we use the corresponding effective area data released by the IceCube
collaboration~\cite{IceCube:2023sov}.
The effective area of a neutrino telescope like IceCube encapsulates the cross-section of a neutrino interacting against the detector target as well as detector-specific effects, including its geometry and efficiencies, the event selection thresholds, and the attenuation of neutrinos while propagating through the Earth.
Event rates are then computed using:
\begin{equation}
  \label{eq:Nevents}
  N_\text{events}
  = T \int \dd{E_\nu} \int \dd{\Omega}\,
  \sum_{\alpha}\left[
  \left\lbrace
  \phi_{\alpha}^{{\text{Earth}}}(E_\nu) +
  \phi_{\bar{\alpha}}^{{\text{Earth}}}(E_\nu)
  \right\rbrace A_\text{eff}^{\alpha + \bar{\alpha}}(E_\nu)
  \right]
  \,,
\end{equation}
where, \( \phi_{{\alpha(\bar{\alpha})}}^{{\text{Earth}}} \) represents the flux of neutrino flavour $\nu_{{\alpha(\bar{\alpha})}}$ on Earth,
\( \phi_{\alpha}^{{\text{Earth}}} 
= \sum_{i=1}^{3} {\abs{U_{\alpha i}}}^2 \phi_{i}^{{\text{Earth}}} \)
and $\phi_{\bar{\alpha}}^{{\text{Earth}}} 
= \sum_{\bar{i}=1}^{3} {\abs{U_{\bar{\alpha} \bar{i}}}}^2 \phi_{\bar{i}}^{{\text{Earth}}}$ (with $U_{\bar{\alpha} \bar{i}} = U^\ast_{\alpha \bar{i}}$),
which are computed from the redshift integrated neutrino flux for the
$i$\textsuperscript{th} mass eigenstate at Earth,
$\phi_{i(\bar{i)}}^{{\text{Earth}}} = \phi_{i(\bar{i)}}(z=0)$, and \( A^{{\alpha + \bar{\alpha}}}_\text{eff} \) corresponds to the flavour specific effective area
summed over particle-antiparticle states.
  The energy dependent effective area \( A_\text{eff}^{{\alpha + \bar{\alpha}}} \) is taken from
  the official IceCube data release~\cite{DVN/PZNO2T_2023} accompanying the HESE
  analysis.
  This includes factors of standard neutrino-nucleon scattering cross-sections
  in ice for both neutral and charged current interactions for each neutrino
  flavour as well as detector-specific efficiency factors.
  For the electron antineutrino component, specifically, it also incorporates
  resonant Glashow scattering process against electrons in the detector medium
  \( \mqty(E_{\bar{\nu}_e} = \SI{6.3}{\PeV}) \).
The total runtime of the experiment is represented by $T$.
For an isotropic flux, we can carry out the angular integral trivially, giving the total number of events:
\begin{equation}
  \label{eq:Nevents-iso}
  N_\text{events}
  = 4 \pi T \int \dd{E_\nu} \,
  \sum_{\alpha}\left[
  \left\lbrace
  \phi_{\alpha}^{{\text{Earth}}}(E_\nu) +
  \phi_{\bar{\alpha}}^{{\text{Earth}}}(E_\nu)
  \right\rbrace A_\text{eff}^{\alpha + \bar{\alpha}}(E_\nu)
  \right]
  \,.
\end{equation}

\begin{figure}[tbh]
  \centering
  \includegraphics[width=0.48\textwidth]{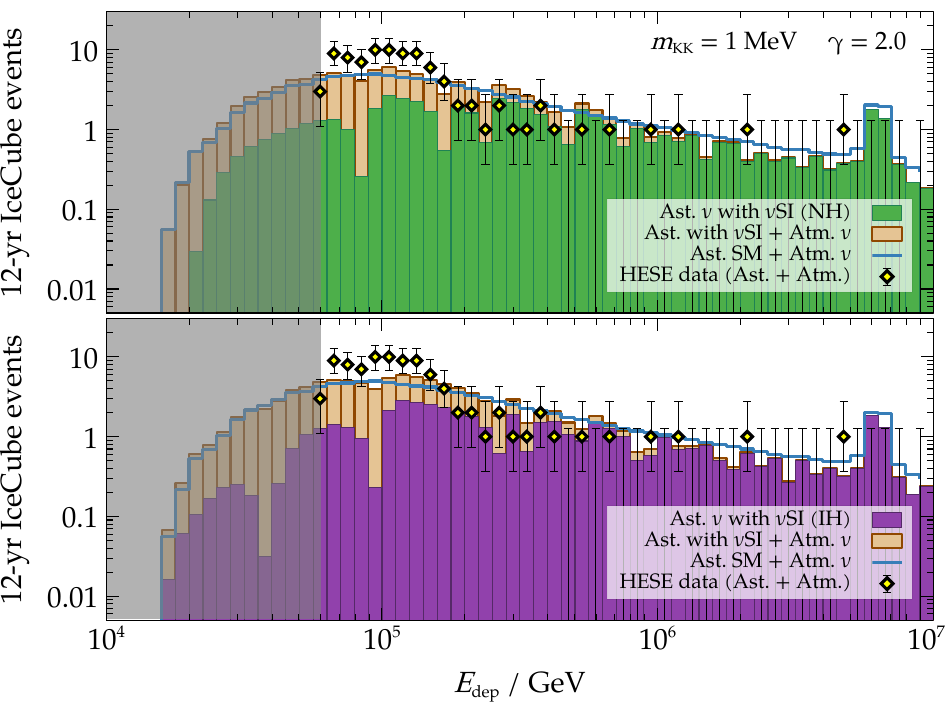}\quad
  \includegraphics[width=0.48\textwidth]{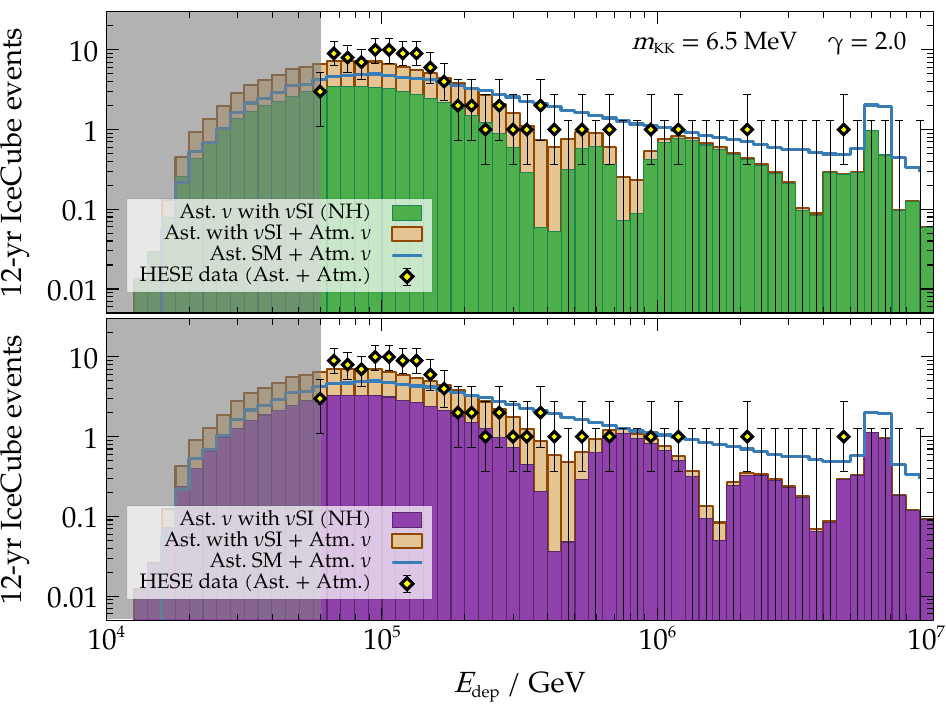}
  \caption{\label{fig:events-mkk10-phi25}
    12-year IceCube HESE numbers
    from self-interaction-induced scattered neutrino fluxes as a function of the
    deposited energy \( E_\text{dep} \).
    An injection spectrum of \( \phi^\text{ini} \propto {E_\nu^{-2}} \) has been
    used along with \(\mkk = \SI{1}{\MeV}\) (left) and \SI{6.5}{\MeV} (right) as
    examples.
    Events for both normal (NH) and inverted hierarchies (IH) of neutrino masses
    are shown.
    We use \( N_\text{KKmax} = 15 \), which is big enough to push the resonances
    corresponding to the highest \zpr\ mode to energies beyond the HESE limits.
    We remind readers that `with $\nu$SI' refers to our 5d scenario.
  }
\end{figure}

In Fig.~\ref{fig:events-mkk10-phi25}, we show 12-year binned HESE estimates
at IceCube corresponding to \( \mkk = \SIlist{1;6.5}{\MeV} \), as representative
cases, using an injection spectrum $\phi^\text{ini} \propto {E_0}^{-2}$ for
both, and considering normal as well as inverted neutrino mass hierarchies.
The assumption of a relatively hard injection flux is made on the basis that the
light \zprn{n}\ mediated attenuation of the flux as it propagates will
soften the high-energy spectrum enough to provide good fits to the sparse
\unit{\PeV} scale events, as well as explaining the more populated bins at the
lower end of the spectrum (\( E_\nu \sim \SI{100}{\TeV} \)).
To determine the overall normalisation of our fluxes $A$, we have normalised the cumulative event numbers thus computed for
\( E_\text{dep} \geqslant \SI{60}{\TeV} \) against the observed HESE signal data
from Ref.~\cite{IceCube:2025cjy}.

For comparison with observed HESE data, we show bin-by-bin events obtained by
summing those from fluxes affected by \zprn{n}\ mediated neutrino
self-interactions ($\nu$SI) (green bars in
Fig.~\ref{fig:events-mkk10-phi25}) with the conventional atmospheric neutrino
background (orange bars stacked atop BSM events).
We also show events predicted for SM neutrino propagation, \ie\
without neutrino self-interactions, for the same selection of spectral index
\(\gamma\), also along with the conventional atmospheric background (blue curve).
This atmospheric neutrino flux is a product of cosmic rays interacting with
  nuclei in the atmosphere to produce light charged mesons
  (\( \pi^{\pm} \) and \( K^{\pm} \)), which lose energy in the atmosphere
  before decaying to produce neutrinos, \( \nu_e \text{ and } \nu_\mu \), and
  their associated charged leptons, \( e^{\pm} \text{ and } \mu^{\pm} \)
  respectively.
We use the flux parametrisation described in Ref.~\cite{Honda:2006qj} and
include both \( \nu_\mu \) and \( \nu_e \) atmospheric fluxes.
On the other hand, the prompt component of atmospheric neutrino fluxes may become
important~\cite{Bhattacharya:2015jpa,Bhattacharya:2016jce,Gauld:2015kvh,Garzelli:2015psa}
at energies exceeding \SI{100}{\TeV}; however, for this analysis, we omit them
on account of the current IceCube studies turning up no corresponding
events~\cite{IceCube:2020wum,IceCube:2023mrq}.

A few features stand out when analysing event rates involving neutrino fluxes
scattered by BSM-mediated interactions during their propagation, as shown in
Fig.~\ref{fig:events-mkk10-phi25}:
\begin{itemize}
  \item Dips in the flux spectrum do translate to the binned event
        rates.\footnote{This inference certainly depends on the choice of bin
          widths. In the current work, we have used bin widths consistent with the effective area data released by IceCube for their 12-year HESE
          data~\cite{DVN/PZNO2T_2023}, which in turn is approximately
          representative of the average energy resolution of events in this
          sampling.}
  \item While there ought to be three dips in the spectrum for each
        \( m_{\zprn{n}} \), corresponding to the three relic neutrino masses,
        those corresponding to neutrino mass eigenstates separated by \( \Delta
        m^2_\text{sol} \) overlap and become indistinguishable.
  \item Despite the assumption of a hard initial spectral flux
        \( \qty(\gamma = 2) \), events from the \( \nu \) self-interaction
        attenuated flux are consistent with a paucity of events seen at most
        bins with energies beyond a \unit{\PeV}.
        This is due to increased attenuation of the UHE neutrino flux at higher
        energies.
  \item As \zprn{n}\ decay widths increase with $n$ [see
            Eq.~\eqref{eq:ed_z_gamma}], corresponding dips in the
        flux/event spectrum become narrower, and for \( n \gtrsim 5\), they
        become impossible to discern individually, given the limited energy
        resolution possible at the ultra-high energies.
        Instead, multiple dips manifest as one conjoined drop-off at the high
        energies.
  \item For low \( \mkk \approx \SI{1}{\MeV} \), dips in the spectrum occur at
        energies near the lower end of the spectrum \( \sim {\SI{100}{\TeV}} \),
        which is in tension with observed data.
  \item Using \( \mkk \approx \SIrange{5}{7}{\MeV} \), produces events that
        are more in tune with the observed HESE data, including spectral dips at
        energies beyond \( \sim \SI{300}{\TeV} \) consistent with several
        null-event bins.
        This includes null-event predictions at the Glashow resonance
        energies, just as reported in the HESE data.
  \item A small excess of events preceding each resonant dip in the spectra is
        a consequence of regeneration of neutrinos from \zprn{n}\ decays [third
        term in the R.H.S.\ of Eq.~\eqref{eq:boltzmann}].
\end{itemize}

\section{Statistical analysis of results}%
\label{sec:statanal}

Having computed event rates for neutrino fluxes attenuated by \zprn{n}\
mediated interactions, we can turn our attention to determining regions of the
parameter spaces that are favoured by fits to observed HESE data.
Overall, the solution of the Boltzmann equations [Eq.~\eqref{eq:boltzmann}] to
determine the nature of neutrino fluxes arriving at Earth hinges on  five main
parameters:
\begin{itemize}
  \item Normalisation of the initial flux, \( A \);
  \item Spectral index assumed for the initial flux, $\gamma$;
  \item Mass of the lightest \zprn{n}\ boson in the \zpr\ tower, $\mkk$;
  \item The coupling constant determining the strength of the
        \( \zprn{n}\nu\nu \) interaction, \( \gpr \); 
  \item The position of the SM brane where all of the SM particles are localised, $y_\text{SM}$.
\end{itemize}

For our analysis, we keep the value of \( \gpr = \num{e-4} \) and $y_\text{SM} = 0$ fixed throughout.
We vary the parameters \( \qty{\mkk, \gamma} \) to determine the resulting flux
and the bin-by-bin events thereof, and adjust the normalisation $A$ to reproduce
the total 12-year HESE number between \SI{60}{\TeV} and \SI{10}{\PeV}, including
astrophysical neutrinos and the conventional atmospheric neutrino background.
By comparing the binned event prediction to observed data, we compute the
$\chi^2 := \chi^2\qty(\mkk, \gamma)$ for each point of the \( 2d \)
parameter space \( \qty{\mkk, \gamma} \).
To determine the contribution of each bin to the \( \chi^2 \), we use:
\[
  \chi^2_{i}(\mkk, \gamma) := \qty[n_i - p_i\qty(\mkk, \gamma)]^2 / \Delta_i^2\,,
\]
where $n_i$ is the observed event number from HESE data corresponding
to the $i$\textsuperscript{th} bin and, likewise, $p_i\qty(\mkk, \gamma)$
are the predicted events at the same bin for a specific $\mkk$ and $\gamma$.
$\Delta_i$ is the symmetrised error obtained using a two-sided 67\% confidence
level Poisson uncertainty around the observed data $n_i$.\footnote{When $n_i =
  0$ at some bin $i$, we use one-sided 67\% confidence level uncertainty for
  $\Delta_i$ instead.
}
The point in the parameter space that produces the minimum value,
\( \chi^2_\text{min} = \chi^2(\mkk^\text{b.f.}, \gamma^\text{b.f.}) \)
defines its best-fit values.
Favoured regions of the parameter space with respect to the best fit point are then inferred from the value of
\[
\Delta\chi^2\qty(\mkk, \gamma) :=
\chi^2\qty(\mkk, \gamma) - \chi^2_\text{min}.
\]

\begin{figure}[tbh]
  \begin{center}
    \includegraphics[scale=0.47]{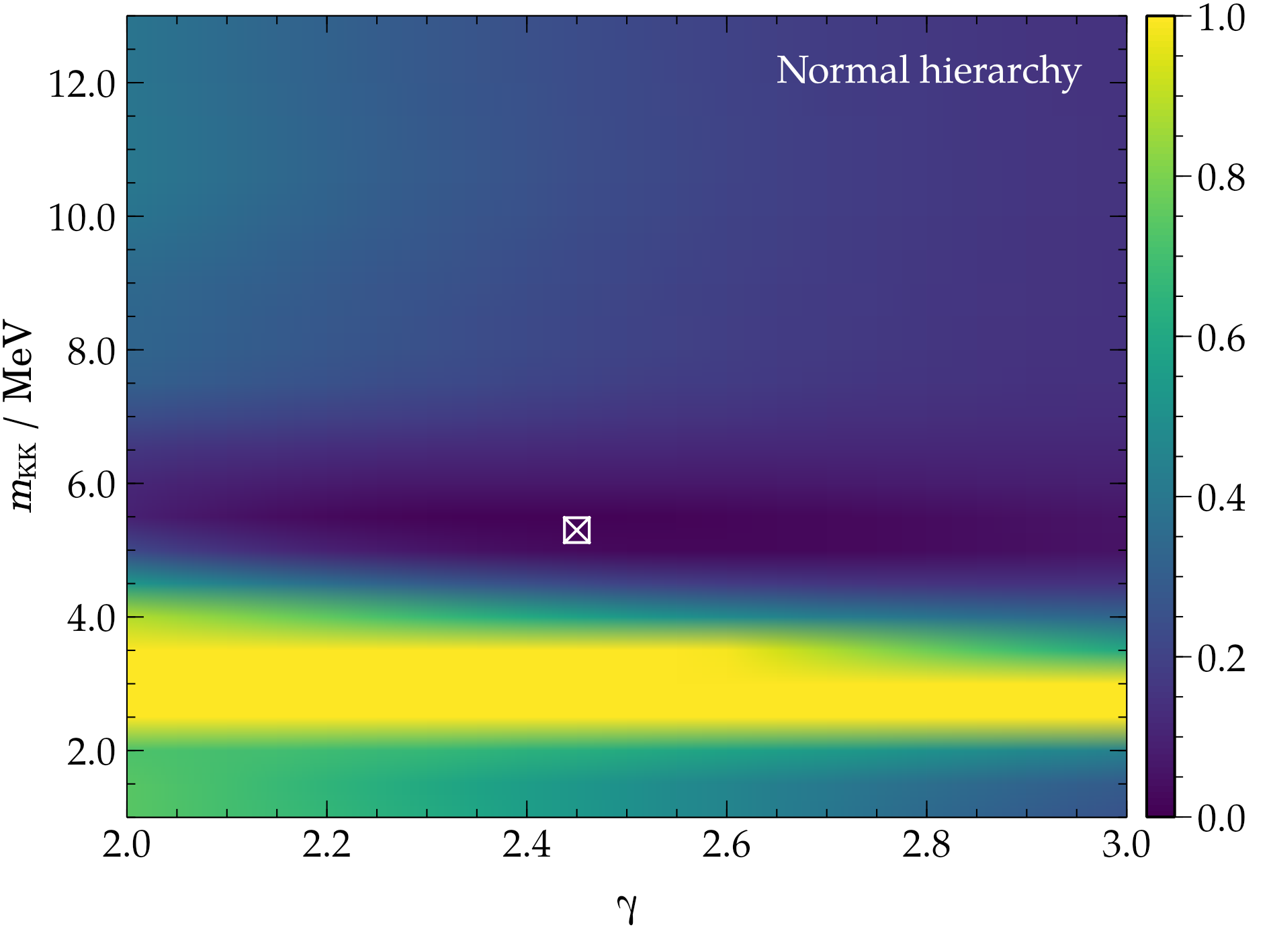}
    \includegraphics[scale=0.47]{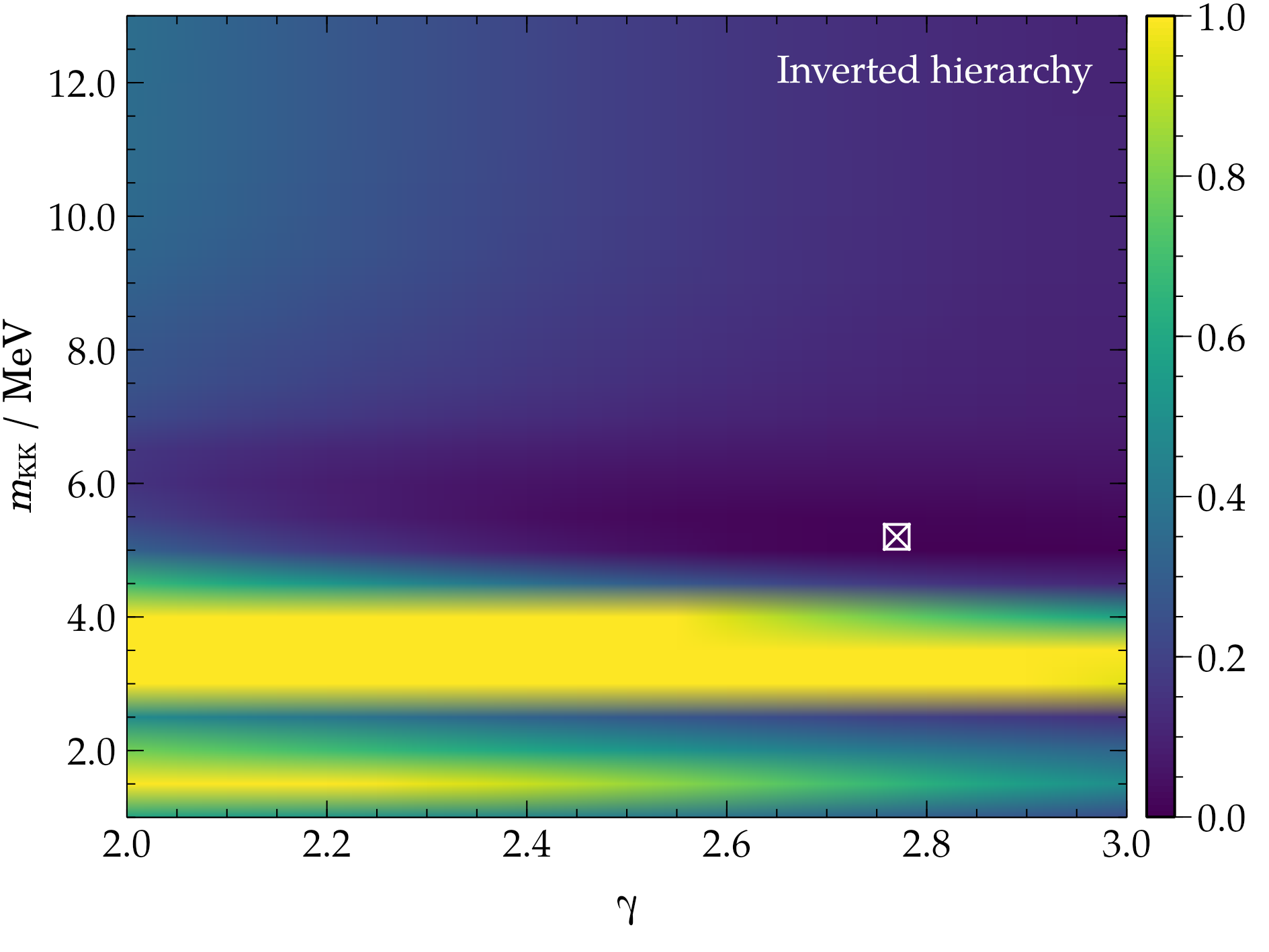}
  \end{center}
  \caption{\label{fig:chisq}Parameter space
    sensitivities for \zprn{n} mediated UHE $\nu$ flux attenuation in the case of
    neutrino masses arranged in a normal hierarchy (left) and an inverted hierarchy
    (right).
    Best-fits in each case are indicated by white crosses, but are largely
    degenerate between \( 2.2 \leq \gamma \leq 2.8 \) for the best-fit value of
    \mkk.
    The colour coding demonstrates the difference between the value of
    \( \chi^{2} \) at any specific point of the
    \( \lbrace m_\text{KK}, \gamma \rbrace \) parameter space and the
    \( \chi^{2}_\text{min} \) at the best-fit:
    \( \Delta\chi^2\qty(\mkk, \gamma)
    = \chi^2\qty(\mkk, \gamma) - \chi^2_\text{min} \).
    The region \( \SI{2.5}{\MeV} \leqslant \mkk \leqslant \SI{4.0}{\MeV} \) is
    disfavoured at more than \( 1\sigma \) level with respect to the best-fit
    for each hierarchy.
  }
\end{figure}

The favoured regions of the \( \qty{\mkk, \gamma} \) parameter space for normal
and inverted neutrino mass hierarchies are shown in Fig.~\ref{fig:chisq}.
We note that current event statistics are not enough to significantly determine
preferred parameter values, and the best-fit parameters obtained by us are
largely degenerate with respect to variations of the spectral index \( \gamma \)
between 2.0 and 3.0.
In general, we find that the \( \chi^2 \) is largely insensitive to the choice
of $\gamma$ for the injection spectra.

On the other hand, for both hierarchies, we find strong sensitivity to \mkk,
which affects the location of resonances in the flux and event spectrum, as long
as \( \mkk \leqslant \SI{7}{\MeV} \).
For \( \mkk \gtrsim \SI{7.5}{\MeV} \), the \( \chi^2 \) becomes nearly
insensitive to \mkk, a consequence of all but the first resonant dip in the flux
spectrum being pushed into \unit{\PeV} energy territory, thereby causing the BSM
flux to largely mirror the SM event rates between
energies of \SI{60}{\TeV} to \SI{1}{\PeV}, where most of the HESE data lie. We also find a more than \( 1\sigma \) preference against the \mkk\ value lying
between \SIrange[range-phrase = { and }]{2.5}{4.0}{\MeV} for both mass
hierarchies.
Conversely, \mkk\ values in the region
\( \SI{5.0}{\MeV} \leqslant \mkk \leqslant \SI{6.5}{\MeV} \) provide good fits
to the data, again largely irrespective of the value of $\gamma$.
However, this must be understood with the caveat that the variation
for the $\chi^2$ across the considered parameter space from its minimum value
is slow, and limited to a maximum value of 1.0.
This is reflective of low signal statistics in the current IceCube HESE data,
and suggests that making model-specific parameter determinations to any significant level may only be possible with the even larger volume future UHE neutrino detector, \gen, or through global analyses incorporating statistics
from multiple detectors, including IceCube, KM3NeT, and others.

Notwithstanding this, the main inference from our results remains that
in comparison to SM fits obtained by IceCube, which find a strong
preference in favour of \( \gamma = 2.74^{+0.06}_{-0.07} \)~\cite{Naab:2023xcz}
in the energy range \( \SI{60}{\TeV} \text{ to } \SI{10}{\PeV} \),
our results show a comparatively weak dependence for \( \gamma \in [2.0, 2.8] \).
In particular, with the introduction of the \zpr\ tower, hard spectral indices for
the initial neutrino flux, such as \( \gamma = 2.0 \) become allowed
whereas they are disfavoured at more than \( 5\sigma \) in the SM only scenario.\footnote{
One likely cause is the large number of zero-event bins (given our choice of bin
size) in the latest data we used. These bins act to reduce the value of $\Delta
\chi^2$, particularly when attenuation is strong, but this may remain to some
extent even when \mkk\ is large, where the situation becomes close to the
SM.}

  We note that, when limiting the statistical analysis to within \SI{10}{\PeV} for
  consistency with HESE data, increasing values of \mkk\ push
  BSM resonances to higher and higher energies, until the effects thereof
  are no longer seen across the specific energy range.
  Even for \( \mkk \sim \SI{10}{\MeV} \), the first resonant peaks
  are pushed into \unit{\PeV} territory, where statistics are sparse, and as a consequence, parameter dependences to fits weaken significantly.
  With even higher \mkk, the fit, insofar as it concerns HESE energies,
  approaches the SM scenario.
  However, given that our BSM model has more parameters than the typical SM single
  power-law fit employed for fits to HESE data, the preference for softer
  $\gamma$ in our analysis, while noticeable, is weak.
  Indeed, a very mild preference away from hard \( \gamma \leqslant 2.4 \) and towards
  softer \( \gamma \approx 2.8 \) is already seen for
  \( \mkk \geqslant \SI{8}{\MeV} \) in Fig.~\ref{fig:nuflux5d} for both mass
  hierarchies.
  
Paper~\cite{Francener:2025apz} employs similar data setup (but a smaller number of zero-event bins) and discusses a 4d $U(1)_{L_\mu - L_\tau}$ model containing a single $Z'$ boson in terms of $N_\text{events}$; in that case, a best-fit $\gamma$ value of approximately $2.8$–$2.9$ is reported (depending on setups); no information is provided regarding standard deviations of $\gamma$ values other than the best-fit.
This value is very close to the result obtained under the SM assumption, indicating that the introduction of a single $Z'$ alone is insufficient to reduce the signal, except at one or two specific energy bins corresponding
to s-channel resonances, and the
value of $\gamma$ remains virtually unchanged.

We do not attempt a statistical analysis to determine whether the BSM-affected scenario provides an overall better fit in comparison to the current
SM power-law best fit with $\gamma \approx 2.9$ because it is
apparent that the dearth of event statistics with deposited
energies \( E_\text{dep} \geqslant \SI{60}{\TeV} \) in current data would
preclude any significant preference for one model or the other.

\section{Discussion and conclusions}%
\label{sec:discon}

Analyses of ultra-high-energy neutrinos, which are produced at extragalactic
astrophysical sources and travel Megaparsec distances thereafter to reach the Earth, provide rich opportunities to investigate BSM-mediated scattering with the intervening medium along their paths.

In this work, we have investigated the consequences of an extra-dimensional
BSM scenario with broken \Umutau\ symmetry, which manifests as a tower of
multiple \unit{\MeV}-scale neutral gauge bosons as part of its particle
spectrum, all of which mediate interactions between the SM \(
\mu\text{-}\tau \) sector.
This includes neutrino self-interactions between \( \nu_\mu \) and
\( \nu_\tau \) flavours.
The existence of multiple mediators capable of inducing scattering processes
between neutrinos, not effective in the SM, naturally leads to a
distinctive cross-section shapes featuring multiple $s$-channel resonant peaks
as well as dips from the interference of different gauge boson-mediated
amplitudes.
The latter may be washed out by $t/u$-channel off-resonant contributions
to the scattering amplitude, and thus, unlike in the single mediator case, may
no longer be ignored in comparison to $s$-channel resonances.
We evaluate the scattering cross-section between astrophysical neutrinos and
relic neutrinos at (lab) energies over \SI{100}{\TeV}, relevant for
UHE neutrinos detected at IceCube.

Starting with the assumption of a single power-law nature for the neutrino flux
produced at source, we solve the Boltzmann equations numerically to understand the flux spectrum of neutrinos arriving at Earth, having been repeatedly
scattered by interactions with the \cnub\ medium via such BSM mediated
processes.
We find that peaks and dips in the cross-section naturally translate to narrow
dips and broad excesses, respectively, in the final neutrino flux spectrum.

We compute event rates corresponding to these fluxes for 12 years of IceCube
runtime using effective areas describing high-energy starting events (HESE),
and find that these features in the flux, modulo energy resolution effects, also
appear in the resulting binned event spectrum.
In addition to the distinctive peaks and dips in the event spectrum at energies
over \SI{100}{\TeV}, we find that, for a range of reasonable values of the model
parameter space, it is also characterised by an attenuation feature that fits
neatly with the paucity of observed events at bins with energies between
\SIrange{1}{10}{\PeV}.
In other words, even if one starts with a hard spectral index for the initial
neutrino flux, repeated interactions of the UHE neutrino with relic neutrinos as the former propagates cosmic distances to Earth, lead to a softening of the final
neutrino flux spectrum reaching Earth.
This physically motivated attenuation may, then, explain the softness of the
flux spectrum observed with energies over \SI{60}{\TeV} at IceCube without
requiring neutrino fluxes to originate with unusually soft spectra.

Having computed the predicted event rates for a range of different values of the
parameter space, we compare these binned event predictions against observed
IceCube 12-year HESE data to determine regions of the parameter space favoured
by the latter.
Due to low statistics at energies above \SI{100}{\TeV}, the preference for any
region of the parameter space remains weak.
However, what emerges is a near-degeneracy with respect to the spectral index
$\gamma$ between the values of \( \numrange{2.0}{2.8} \), for \mkk\ ranging
between \qtyrange{5}{10}{\MeV}, that is distinct from
the statistically significant preference for soft spectra \( \gamma \approx 2.85 \)
obtained in the SM scenario using single power-law fits.

Additionally, our results demonstrate features in the $\nu$ event spectrum that future UHE neutrino detectors will be able to discern.
Courtesy of the distinctive, recurring dip and rise features in the flux spectrum
that are a consequence of multiple \zprn{n}\ mediated interactions of
astrophysical neutrinos with relic neutrinos in the \cnub, the extra-dimensional model involving a broken \Umutau\ gauge symmetry analysed in this work lends
itself to testability at large volume neutrino detectors like IceCube with more
statistics --- and KM3NeT, \gen, and others in the future.

\section*{Acknowledgments}
AK thanks Dr.~Kenny Ng for useful discussions related to the project.
AB thanks Prof. Ina Sarcevic and Prof. Jean-René Cudell for useful discussions.
The authors acknowledge the Shiv Nadar Institution of Eminence for providing
the computational infrastructure to support this work.

\bibliographystyle{JHEP}
\bibliography{refs}

@article{Escudero:2019gzq,
    author = "Escudero, Miguel and Hooper, Dan and Krnjaic, Gordan and Pierre, Mathias",
    title = "{Cosmology with A Very Light ${L_{\mu} - L_{\tau}} $ Gauge Boson}",
    eprint = "1901.02010",
    archivePrefix = "arXiv",
    primaryClass = "hep-ph",
    reportNumber = "FERMILAB-PUB-19-001-A, LPT-Orsay-18-15, IFIC-19-02, KCL-19-01,
  IFT-UAM/CSIC-19-7, KCL-19-01",
    doi = "10.1007/JHEP03(2019)071",
    journal = "JHEP",
    volume = "03",
    pages = "071",
    year = "2019"
}

@article{Araki:2021xdk,
    author = "Araki, Takeshi and Asai, Kento and Honda, Kei and Kasuya, Ryuta and Sato, Joe and Shimomura, Takashi and Yang, Masaki J. S.",
    title = "{Resolving the Hubble tension in a $U(1)_{L_\mu-L_\tau}$ model with the Majoron}",
    eprint = "2103.07167",
    archivePrefix = "arXiv",
    primaryClass = "hep-ph",
    reportNumber = "STUPP-20-242, UME-PP-18",
    doi = "10.1093/ptep/ptab108",
    journal = "PTEP",
    volume = "2021",
    number = "10",
    pages = "103B05",
    year = "2021"
}

@article{Carpio:2021jhu,
    author = "Carpio, Jose Alonso and Murase, Kohta and Shoemaker, Ian M. and Tabrizi, Zahra",
    title = "{High-energy cosmic neutrinos as a probe of the vector mediator scenario in light of the muon $g-2$ anomaly and Hubble tension}",
    eprint = "2104.15136",
    archivePrefix = "arXiv",
    primaryClass = "hep-ph",
    doi = "10.1103/PhysRevD.107.103057",
    journal = "Phys. Rev. D",
    volume = "107",
    number = "10",
    pages = "103057",
    year = "2023"
}

@article{Asai:2023ajh,
    author = "Asai, Kento and Asano, Tomoya and Sato, Joe and Yang, Masaki J. S.",
    title = "{Contribution of Majoron to Hubble tension in gauged $U(1)_{L_\mu-L_\tau}$ Model}",
    eprint = "2309.01162",
    archivePrefix = "arXiv",
    primaryClass = "hep-ph",
    reportNumber = "STUPP-23-263",
    month = "9",
    year = "2023"
}

@article{Kayser:2004wmw,
    author = "Kayser, Boris",
    editor = "Hewett, Joanne and Jaros, John and Kamae, Tsuneyoshi and Prescott, Charles",
    title = "{Neutrino physics}",
    eprint = "hep-ph/0506165",
    archivePrefix = "arXiv",
    reportNumber = "SSI-2004-L004, FERMILAB-PUB-05-236-T",
    journal = "eConf",
    volume = "C040802",
    pages = "L004",
    year = "2004"
}

@article{IceCube:2023mrq,
    author = "Boettcher, Jakob and others",
    collaboration = "IceCube",
    title = "{Search for the Prompt Atmospheric Neutrino Flux in IceCube}",
    eprint = "2309.07560",
    archivePrefix = "arXiv",
    primaryClass = "astro-ph.HE",
    reportNumber = "PoS-ICRC2023-1068",
    doi = "10.22323/1.444.1068",
    journal = "PoS",
    volume = "ICRC2023",
    pages = "1068",
    year = "2023"
}

@article{Francener:2025apz,
    author = "Francener, Reinaldo and Goncalves, Victor P. and Gratieri, Diego R.",
    title = "{Probing a low-mass Z' gauge boson at IceCube and prospects for IceCube-Gen2}",
    eprint = "2502.19338",
    archivePrefix = "arXiv",
    primaryClass = "hep-ph",
    doi = "10.1103/PhysRevD.111.095005",
    journal = "Phys. Rev. D",
    volume = "111",
    number = "9",
    pages = "095005",
    year = "2025"
}

@article{Ibe:2025rwk,
    author = "Ibe, Masahiro and Shirai, Satoshi and Watanabe, Keiichi",
    title = "{Global neutrino constraints on the minimal U(1)L{\ensuremath{\mu}}-L{\ensuremath{\tau}} model}",
    eprint = "2503.01399",
    archivePrefix = "arXiv",
    primaryClass = "hep-ph",
    reportNumber = "IPMU25-0011",
    doi = "10.1103/PhysRevD.111.095034",
    journal = "Phys. Rev. D",
    volume = "111",
    number = "9",
    pages = "095034",
    year = "2025"
}

@article{KM3NeT:2025npi,
    author = "Aiello, S. and others",
    collaboration = "KM3NeT",
    title = "{Observation of an ultra-high-energy cosmic neutrino with KM3NeT}",
    doi = "10.1038/s41586-024-08543-1",
    journal = "Nature",
    volume = "638",
    number = "8050",
    pages = "376--382",
    year = "2025",
    note = "[Erratum: Nature 640, E3 (2025)]"
}

@article{KM3Net:2016zxf,
    author = "Adrian-Martinez, S. and others",
    collaboration = "KM3Net",
    title = "{Letter of intent for KM3NeT 2.0}",
    eprint = "1601.07459",
    archivePrefix = "arXiv",
    primaryClass = "astro-ph.IM",
    doi = "10.1088/0954-3899/43/8/084001",
    journal = "J. Phys. G",
    volume = "43",
    number = "8",
    pages = "084001",
    year = "2016"
}

@article{DiFranzo:2015qea,
    author = "DiFranzo, Anthony and Hooper, Dan",
    title = "{Searching for MeV-Scale Gauge Bosons with IceCube}",
    eprint = "1507.03015",
    archivePrefix = "arXiv",
    primaryClass = "hep-ph",
    reportNumber = "FERMILAB-PUB-15-292-A, UCI-HEP-TR-2015-11",
    doi = "10.1103/PhysRevD.92.095007",
    journal = "Phys. Rev. D",
    volume = "92",
    number = "9",
    pages = "095007",
    year = "2015"
}

@article{Esteban:2021tub,
    author = "Esteban, Ivan and Pandey, Sujata and Brdar, Vedran and Beacom, John F.",
    title = "{Probing secret interactions of astrophysical neutrinos in the high-statistics era}",
    eprint = "2107.13568",
    archivePrefix = "arXiv",
    primaryClass = "hep-ph",
    reportNumber = "FERMILAB-PUB-21-328-T, nuhep-th/21-06",
    doi = "10.1103/PhysRevD.104.123014",
    journal = "Phys. Rev. D",
    volume = "104",
    number = "12",
    pages = "123014",
    year = "2021"
}

@article{IceCube:2025cjy,
    author = "Lad, Neha Navnitkumar and others",
    collaboration = "IceCube",
    title = "{Neutrino flavor composition using High Energy Starting Events with IceCube}",
    eprint = "2507.06835",
    archivePrefix = "arXiv",
    primaryClass = "astro-ph.HE",
    reportNumber = "PoS-ICRC2025-1198",
    doi = "10.22323/1.501.1198",
    journal = "PoS",
    volume = "ICRC2025",
    pages = "1198",
    year = "2025"
}

@article{Hooper:2023fqn,
    author = "Hooper, Dan and Iguaz Juan, Joaquim and Serpico, Pasquale D.",
    title = "{Signals of a new gauge boson from IceCube and the muon g-2}",
    eprint = "2302.03571",
    archivePrefix = "arXiv",
    primaryClass = "astro-ph.HE",
    reportNumber = "FERMILAB-PUB-23-038-T",
    doi = "10.1103/PhysRevD.108.023007",
    journal = "Phys. Rev. D",
    volume = "108",
    number = "2",
    pages = "023007",
    year = "2023"
}

@article{Ng:2014pca,
    author = "Ng, Kenny C. Y. and Beacom, John F.",
    title = "{Cosmic neutrino cascades from secret neutrino interactions}",
    eprint = "1404.2288",
    archivePrefix = "arXiv",
    primaryClass = "astro-ph.HE",
    doi = "10.1103/PhysRevD.90.065035",
    journal = "Phys. Rev. D",
    volume = "90",
    number = "6",
    pages = "065035",
    year = "2014",
    note = "[Erratum: Phys.Rev.D 90, 089904 (2014)]"
}

@article{Planck:2018vyg,
    author = "Aghanim, N. and others",
    collaboration = "Planck",
    title = "{Planck 2018 results. VI. Cosmological parameters}",
    eprint = "1807.06209",
    archivePrefix = "arXiv",
    primaryClass = "astro-ph.CO",
    doi = "10.1051/0004-6361/201833910",
    journal = "Astron. Astrophys.",
    volume = "641",
    pages = "A6",
    year = "2020",
    note = "[Erratum: Astron.Astrophys. 652, C4 (2021)]"
}

@article{Yuksel:2008cu,
    author = "Yuksel, Hasan and Kistler, Matthew D. and Beacom, John F. and Hopkins, Andrew M.",
    title = "{Revealing the High-Redshift Star Formation Rate with Gamma-Ray Bursts}",
    eprint = "0804.4008",
    archivePrefix = "arXiv",
    primaryClass = "astro-ph",
    doi = "10.1086/591449",
    journal = "Astrophys. J. Lett.",
    volume = "683",
    pages = "L5--L8",
    year = "2008"
}

@article{Francener:2024bfm,
    author = "Francener, Reinaldo and Goncalves, Victor P. and Gratieri, Diego R.",
    title = "{Sensitivity of the neutrino transmission coefficient at high energies to the Earth{\textquoteright}s density profile}",
    eprint = "2403.16611",
    archivePrefix = "arXiv",
    primaryClass = "hep-ph",
    doi = "10.1088/1361-6471/ade4e4",
    journal = "J. Phys. G",
    volume = "52",
    number = "7",
    pages = "075201",
    year = "2025"
}

@article{Abbasi:2021qfz,
    author = "Abbasi, R. and others",
    title = "{Improved Characterization of the Astrophysical Muon{\textendash}neutrino Flux with 9.5 Years of IceCube Data}",
    eprint = "2111.10299",
    archivePrefix = "arXiv",
    primaryClass = "astro-ph.HE",
    doi = "10.3847/1538-4357/ac4d29",
    journal = "Astrophys. J.",
    volume = "928",
    number = "1",
    pages = "50",
    year = "2022"
}

@article{Kamada:2015era,
    author = "Kamada, Ayuki and Yu, Hai-Bo",
    title = "{Coherent Propagation of PeV Neutrinos and the Dip in the Neutrino Spectrum at IceCube}",
    eprint = "1504.00711",
    archivePrefix = "arXiv",
    primaryClass = "hep-ph",
    doi = "10.1103/PhysRevD.92.113004",
    journal = "Phys. Rev. D",
    volume = "92",
    number = "11",
    pages = "113004",
    year = "2015"
}

@article{IceCube:2023sov,
    author = "Abbasi, Rasha and others",
    collaboration = "IceCube",
    title = "{Updated directions of IceCube HESE events with the latest ice model using DirectFit}",
    eprint = "2307.13878",
    archivePrefix = "arXiv",
    primaryClass = "astro-ph.HE",
    reportNumber = "PoS-ICRC2023-1030, PoS-ICRC-1030",
    doi = "10.22323/1.444.1030",
    journal = "PoS",
    volume = "ICRC2023",
    pages = "1030",
    year = "2023"
}

@article{IceCube:2025tgp,
    author = "Abbasi, R. and others",
    collaboration = "IceCube",
    title = "{Evidence for a Spectral Break or Curvature in the Spectrum of Astrophysical Neutrinos from 5 TeV--10 PeV}",
    eprint = "2507.22233",
    archivePrefix = "arXiv",
    primaryClass = "astro-ph.HE",
    doi = "10.1103/2gh9-d4q7",
    journal = "Phys. Rev. Lett.",
    volume = "136",
    pages = "121002",
    year = "2026"
}

@article{IceCube:2020wum,
    author = "Abbasi, R. and others",
    collaboration = "IceCube",
    title = "{The IceCube high-energy starting event sample: Description and flux characterization with 7.5 years of data}",
    eprint = "2011.03545",
    archivePrefix = "arXiv",
    primaryClass = "astro-ph.HE",
    doi = "10.1103/PhysRevD.104.022002",
    journal = "Phys. Rev. D",
    volume = "104",
    pages = "022002",
    year = "2021"
}

@article{Wang:2025qap,
    author = "Wang, Isaac R. and Xu, Xun-Jie and Zhou, Bei",
    title = "{Widen the Resonance: Probing a New Regime of Neutrino Self-Interactions with Astrophysical Neutrinos}",
    eprint = "2501.07624",
    archivePrefix = "arXiv",
    primaryClass = "hep-ph",
    reportNumber = "FERMILAB-PUB-25-0016-T",
    doi = "10.1103/9ddp-j1z9",
    journal = "Phys. Rev. Lett.",
    volume = "135",
    number = "18",
    pages = "181002",
    year = "2025"
}

@article{Machado:2025ltu,
    author = "Machado, Pedro A. N. and Wang, Isaac R. and Xu, Xun-Jie and Zhou, Bei",
    title = "{Widen the Resonance at Ultra-High Energies: Novel Probes of Neutrino Self-interactions in the High-Mass Regime}",
    eprint = "2512.00165",
    archivePrefix = "arXiv",
    primaryClass = "hep-ph",
    reportNumber = "FERMILAB-PUB-25-0853-T",
    month = "11",
    year = "2025"
}

@article{Esteban:2024eli,
    author = "Esteban, Ivan and Gonzalez-Garcia, M. C. and Maltoni, Michele and Martinez-Soler, Ivan and Pinheiro, Jo{\~a}o Paulo and Schwetz, Thomas",
    title = "{NuFit-6.0: updated global analysis of three-flavor neutrino oscillations}",
    eprint = "2410.05380",
    archivePrefix = "arXiv",
    primaryClass = "hep-ph",
    reportNumber = "IFT-UAM/CSIC-24-140, YITP-SB-2024-24, IPPP/24/64, IPPP/24/64, IFT-UAM/CSIC-24-140, YITP-SB-2024-24",
    doi = "10.1007/JHEP12(2024)216",
    journal = "JHEP",
    volume = "12",
    pages = "216",
    year = "2024"
}

@article{ParticleDataGroup:2026aaa,
    author = "Takahashi, F. and others",
    collaboration = "Particle Data Group",
    title = "{Review of Particle Physics}",
    doi = "10.1142/S0217751X26300115",
    journal = "Int. J. Mod. Phys. A",
    volume = "41",
    pages = "2630011",
    year = "2026"
}

@article{IceCube:2021rpz,
    author = "Aartsen, M. G. and others",
    collaboration = "IceCube",
    title = "{Detection of a particle shower at the Glashow resonance with IceCube}",
    eprint = "2110.15051",
    archivePrefix = "arXiv",
    primaryClass = "hep-ex",
    doi = "10.1038/s41586-021-03256-1",
    journal = "Nature",
    volume = "591",
    number = "7849",
    pages = "220--224",
    year = "2021",
    note = "[Erratum: Nature 592, E11 (2021)]"
}

@article{Bhattacharya:2011qu,
    author = "Bhattacharya, Atri and Gandhi, Raj and Rodejohann, Werner and Watanabe, Atsushi",
    title = "{The Glashow resonance at IceCube: signatures, event rates and $pp$ vs. $p\gamma$ interactions}",
    eprint = "1108.3163",
    archivePrefix = "arXiv",
    primaryClass = "astro-ph.HE",
    doi = "10.1088/1475-7516/2011/10/017",
    journal = "JCAP",
    volume = "10",
    pages = "017",
    year = "2011"
}

@article{Muzio:2021zud,
    author = "Muzio, Marco Stein and Farrar, Glennys R. and Unger, Michael",
    title = "{Probing the environments surrounding ultrahigh energy cosmic ray accelerators and their implications for astrophysical neutrinos}",
    eprint = "2108.05512",
    archivePrefix = "arXiv",
    primaryClass = "astro-ph.HE",
    doi = "10.1103/PhysRevD.105.023022",
    journal = "Phys. Rev. D",
    volume = "105",
    number = "2",
    pages = "023022",
    year = "2022"
}

@article{Bhattacharya:2023mmp,
    author = "Bhattacharya, Atri and Enberg, Rikard and Reno, Mary Hall and Sarcevic, Ina",
    title = "{Energy-dependent flavour ratios in neutrino telescopes from charm}",
    eprint = "2309.09139",
    archivePrefix = "arXiv",
    primaryClass = "astro-ph.HE",
    doi = "10.1088/1475-7516/2024/03/057",
    journal = "JCAP",
    volume = "03",
    pages = "057",
    year = "2024"
}

@article{IceCube:2024fxo,
    author = "Abbasi, R. and others",
    collaboration = "IceCube",
    title = "{Characterization of the astrophysical diffuse neutrino flux using starting track events in IceCube}",
    eprint = "2402.18026",
    archivePrefix = "arXiv",
    primaryClass = "astro-ph.HE",
    doi = "10.1103/PhysRevD.110.022001",
    journal = "Phys. Rev. D",
    volume = "110",
    number = "2",
    pages = "022001",
    year = "2024"
}

@misc{DVN/PZNO2T_2023,
author = {IceCube Collaboration},
publisher = {Harvard Dataverse},
title = {{IceCube HESE 12-year data release}},
UNF = {UNF:6:kBwkHCWHVSu1JKMfVaTtdA==},
year = {2023},
version = {V2},
doi = {10.7910/DVN/PZNO2T},
url = {https://doi.org/10.7910/DVN/PZNO2T}
}

@article{Honda:2006qj,
    author = "Honda, Morihiro and Kajita, T. and Kasahara, K. and Midorikawa, S. and Sanuki, T.",
    title = "{Calculation of atmospheric neutrino flux using the interaction model calibrated with atmospheric muon data}",
    eprint = "astro-ph/0611418",
    archivePrefix = "arXiv",
    doi = "10.1103/PhysRevD.75.043006",
    journal = "Phys. Rev. D",
    volume = "75",
    pages = "043006",
    year = "2007"
}

@article{Bhattacharya:2015jpa,
    author = "Bhattacharya, Atri and Enberg, Rikard and Reno, Mary Hall and Sarcevic, Ina and Stasto, Anna",
    title = "{Perturbative charm production and the prompt atmospheric neutrino flux in light of RHIC and LHC}",
    eprint = "1502.01076",
    archivePrefix = "arXiv",
    primaryClass = "hep-ph",
    reportNumber = "NORDITA-2015-9",
    doi = "10.1007/JHEP06(2015)110",
    journal = "JHEP",
    volume = "06",
    pages = "110",
    year = "2015"
}

@article{Bhattacharya:2016jce,
    author = "Bhattacharya, Atri and Enberg, Rikard and Jeong, Yu Seon and Kim, C. S. and Reno, Mary Hall and Sarcevic, Ina and Stasto, Anna",
    title = "{Prompt atmospheric neutrino fluxes: perturbative QCD models and nuclear effects}",
    eprint = "1607.00193",
    archivePrefix = "arXiv",
    primaryClass = "hep-ph",
    doi = "10.1007/JHEP11(2016)167",
    journal = "JHEP",
    volume = "11",
    pages = "167",
    year = "2016"
}

@article{Garzelli:2015psa,
    author = "Garzelli, M. V. and Moch, S. and Sigl, G.",
    title = "{Lepton fluxes from atmospheric charm revisited}",
    eprint = "1507.01570",
    archivePrefix = "arXiv",
    primaryClass = "hep-ph",
    reportNumber = "DESY-15-107, MITP-15-049",
    doi = "10.1007/JHEP10(2015)115",
    journal = "JHEP",
    volume = "10",
    pages = "115",
    year = "2015"
}

@article{Gauld:2015kvh,
    author = "Gauld, Rhorry and Rojo, Juan and Rottoli, Luca and Sarkar, Subir and Talbert, Jim",
    title = "{The prompt atmospheric neutrino flux in the light of LHCb}",
    eprint = "1511.06346",
    archivePrefix = "arXiv",
    primaryClass = "hep-ph",
    reportNumber = "OUTP-15-26P, DCPT-15-128, IPPP-15-64",
    doi = "10.1007/JHEP02(2016)130",
    journal = "JHEP",
    volume = "02",
    pages = "130",
    year = "2016"
}

@article{Skrzypek:2025tmg,
    author = "Skrzypek, Barbara and Klein, Spencer",
    title = "{Measuring the Astrophysical Neutrino-Antineutrino Ratio with IceCube}",
    eprint = "2508.02795",
    archivePrefix = "arXiv",
    primaryClass = "astro-ph.HE",
    doi = "10.22323/1.501.1181",
    journal = "PoS",
    volume = "ICRC2025",
    pages = "1181",
    year = "2025"
}

@article{Foot:1990mn,
    author = "Foot, Robert",
    title = "{New Physics From Electric Charge Quantization?}",
    reportNumber = "MAD/TH/90-14",
    doi = "10.1142/S0217732391000543",
    journal = "Mod. Phys. Lett. A",
    volume = "6",
    pages = "527--530",
    year = "1991"
}

@article{He:1990pn,
    author = "He, X. G. and Joshi, Girish C. and Lew, H. and Volkas, R. R.",
    title = "{NEW Z-prime PHENOMENOLOGY}",
    reportNumber = "UM-P-90/42, OZ-P-90/16",
    doi = "10.1103/PhysRevD.43.R22",
    journal = "Phys. Rev. D",
    volume = "43",
    pages = "22--24",
    year = "1991"
}

@article{He:1991qd,
    author = "He, Xiao-Gang and Joshi, Girish C. and Lew, H. and Volkas, R. R.",
    title = "{Simplest Z-prime model}",
    reportNumber = "CERN-TH-6084-91, UM-P-91-32, OZ-91-07",
    doi = "10.1103/PhysRevD.44.2118",
    journal = "Phys. Rev. D",
    volume = "44",
    pages = "2118--2132",
    year = "1991"
}

@article{Foot:1994vd,
    author = "Foot, Robert and He, X. G. and Lew, H. and Volkas, R. R.",
    title = "{Model for a light Z-prime boson}",
    eprint = "hep-ph/9401250",
    archivePrefix = "arXiv",
    reportNumber = "OITS-532, UM-P-93-115, OZ-93-26, IP-ASTP-32",
    doi = "10.1103/PhysRevD.50.4571",
    journal = "Phys. Rev. D",
    volume = "50",
    pages = "4571--4580",
    year = "1994"
}

@article{Asai:2017ryy,
    author = "Asai, Kento and Hamaguchi, Koichi and Nagata, Natsumi",
    title = "{Predictions for the neutrino parameters in the minimal gauged U(1)$_{L_\mu-L_\tau}$ model}",
    eprint = "1705.00419",
    archivePrefix = "arXiv",
    primaryClass = "hep-ph",
    reportNumber = "UT-17-17, IPMU-17-0072",
    doi = "10.1140/epjc/s10052-017-5348-x",
    journal = "Eur. Phys. J. C",
    volume = "77",
    number = "11",
    pages = "763",
    year = "2017"
}

@article{Asai:2018ocx,
    author = "Asai, Kento and Hamaguchi, Koichi and Nagata, Natsumi and Tseng, Shih-Yen and Tsumura, Koji",
    title = "{Minimal Gauged U(1)$_{L_\alpha - L_\beta}$ Models Driven into a Corner}",
    eprint = "1811.07571",
    archivePrefix = "arXiv",
    primaryClass = "hep-ph",
    reportNumber = "UT-18-28, IPMU18-0191, KUNS-2740",
    doi = "10.1103/PhysRevD.99.055029",
    journal = "Phys. Rev. D",
    volume = "99",
    number = "5",
    pages = "055029",
    year = "2019"
}

@article{Asai:2020qlp,
    author = "Asai, Kento and Okawa, Shohei and Tsumura, Koji",
    title = "{Search for $ \mathrm{U}{(1)}_{L_{\mu }-{L}_{\tau }} $ charged dark matter with neutrino telescope}",
    eprint = "2011.03165",
    archivePrefix = "arXiv",
    primaryClass = "hep-ph",
    reportNumber = "KYUSHU-HET-217",
    doi = "10.1007/JHEP03(2021)047",
    journal = "JHEP",
    volume = "03",
    pages = "047",
    year = "2021"
}

@article{Halzen:2010yj,
    author = "Halzen, Francis and Klein, Spencer R.",
    title = "{IceCube: An Instrument for Neutrino Astronomy}",
    eprint = "1007.1247",
    archivePrefix = "arXiv",
    primaryClass = "astro-ph.HE",
    doi = "10.1063/1.3480478",
    journal = "Rev. Sci. Instrum.",
    volume = "81",
    pages = "081101",
    year = "2010"
}

@article{IceCube:2013cdw,
    author = "Aartsen, M. G. and others",
    collaboration = "IceCube",
    title = "{First observation of PeV-energy neutrinos with IceCube}",
    eprint = "1304.5356",
    archivePrefix = "arXiv",
    primaryClass = "astro-ph.HE",
    doi = "10.1103/PhysRevLett.111.021103",
    journal = "Phys. Rev. Lett.",
    volume = "111",
    pages = "021103",
    year = "2013"
}

@article{IceCube:2013low,
    author = "Aartsen, M. G. and others",
    collaboration = "IceCube",
    title = "{Evidence for High-Energy Extraterrestrial Neutrinos at the IceCube Detector}",
    eprint = "1311.5238",
    archivePrefix = "arXiv",
    primaryClass = "astro-ph.HE",
    doi = "10.1126/science.1242856",
    journal = "Science",
    volume = "342",
    pages = "1242856",
    year = "2013"
}

@article{IceCube-Gen2:2020qha,
    author = "Aartsen, M. G. and others",
    collaboration = "IceCube-Gen2",
    title = "{IceCube-Gen2: the window to the extreme Universe}",
    eprint = "2008.04323",
    archivePrefix = "arXiv",
    primaryClass = "astro-ph.HE",
    doi = "10.1088/1361-6471/abbd48",
    journal = "J. Phys. G",
    volume = "48",
    number = "6",
    pages = "060501",
    year = "2021"
}

@article{Waxman:1997ti,
    author = "Waxman, Eli and Bahcall, John N.",
    title = "{High-energy neutrinos from cosmological gamma-ray burst fireballs}",
    eprint = "astro-ph/9701231",
    archivePrefix = "arXiv",
    reportNumber = "IASSNS-AST-97-14",
    doi = "10.1103/PhysRevLett.78.2292",
    journal = "Phys. Rev. Lett.",
    volume = "78",
    pages = "2292--2295",
    year = "1997"
}

@article{Waxman:1998yy,
    author = "Waxman, Eli and Bahcall, John N.",
    title = "{High-energy neutrinos from astrophysical sources: An Upper bound}",
    eprint = "hep-ph/9807282",
    archivePrefix = "arXiv",
    reportNumber = "IASSNS-AST-98-38",
    doi = "10.1103/PhysRevD.59.023002",
    journal = "Phys. Rev. D",
    volume = "59",
    pages = "023002",
    year = "1999"
}

@article{Bhattacharjee:1999mup,
    author = "Bhattacharjee, Pijushpani and Sigl, Gunter",
    title = "{Origin and propagation of extremely high-energy cosmic rays}",
    eprint = "astro-ph/9811011",
    archivePrefix = "arXiv",
    doi = "10.1016/S0370-1573(99)00101-5",
    journal = "Phys. Rept.",
    volume = "327",
    pages = "109--247",
    year = "2000"
}

@article{Piran:2004ba,
    author = "Piran, Tsvi",
    title = "{The physics of gamma-ray bursts}",
    eprint = "astro-ph/0405503",
    archivePrefix = "arXiv",
    doi = "10.1103/RevModPhys.76.1143",
    journal = "Rev. Mod. Phys.",
    volume = "76",
    pages = "1143--1210",
    year = "2004"
}

@article{Berezinsky:2002nc,
    author = "Berezinsky, V. and Gazizov, A. Z. and Grigorieva, S. I.",
    title = "{On astrophysical solution to ultrahigh-energy cosmic rays}",
    eprint = "hep-ph/0204357",
    archivePrefix = "arXiv",
    doi = "10.1103/PhysRevD.74.043005",
    journal = "Phys. Rev. D",
    volume = "74",
    pages = "043005",
    year = "2006"
}

@article{Blandford:1987pw,
    author = "Blandford, R. and Eichler, D.",
    title = "{Particle Acceleration at Astrophysical Shocks: A Theory of Cosmic Ray Origin}",
    doi = "10.1016/0370-1573(87)90134-7",
    journal = "Phys. Rept.",
    volume = "154",
    pages = "1--75",
    year = "1987"
}

@article{Halzen:2002pg,
    author = "Halzen, Francis and Hooper, Dan",
    title = "{High-energy neutrino astronomy: The Cosmic ray connection}",
    eprint = "astro-ph/0204527",
    archivePrefix = "arXiv",
    doi = "10.1088/0034-4885/65/7/201",
    journal = "Rept. Prog. Phys.",
    volume = "65",
    pages = "1025--1078",
    year = "2002"
}

@article{Kachelriess:2019oqu,
    author = "Kachelriess, M. and Semikoz, D. V.",
    title = "{Cosmic Ray Models}",
    eprint = "1904.08160",
    archivePrefix = "arXiv",
    primaryClass = "astro-ph.HE",
    doi = "10.1016/j.ppnp.2019.07.002",
    journal = "Prog. Part. Nucl. Phys.",
    volume = "109",
    pages = "103710",
    year = "2019"
}

@article{Fermi:1949ee,
    author = "Fermi, Enrico",
    title = "{On the Origin of the Cosmic Radiation}",
    doi = "10.1103/PhysRev.75.1169",
    journal = "Phys. Rev.",
    volume = "75",
    pages = "1169--1174",
    year = "1949"
}

@article{Aliberti:2025beg,
    author = "Aliberti, R. and others",
    title = "{The anomalous magnetic moment of the muon in the Standard Model: an update}",
    eprint = "2505.21476",
    archivePrefix = "arXiv",
    primaryClass = "hep-ph",
    reportNumber = "CERN-TH-2025-101, FERMILAB-PUB-25-0344-T, INT-PUB-25-015, IPARCOS-UCM-25-029, KEK Preprint 2025-22, LTH 1403, MITP-25-037, UWThPh 2025-15, UWThPh
  2025-15, ZU-TH 37/25, IPARCOS-UCM-25-029",
    doi = "10.1016/j.physrep.2025.08.002",
    journal = "Phys. Rept.",
    volume = "1143",
    pages = "1--158",
    year = "2025"
}

@article{Baek:2001kca,
    author = "Baek, Seungwon and Deshpande, N. G. and He, X. G. and Ko, P.",
    title = "{Muon anomalous g-2 and gauged L(muon) - L(tau) models}",
    eprint = "hep-ph/0104141",
    archivePrefix = "arXiv",
    reportNumber = "KAIST-TH-2001-08",
    doi = "10.1103/PhysRevD.64.055006",
    journal = "Phys. Rev. D",
    volume = "64",
    pages = "055006",
    year = "2001"
}

@article{Chakraborty:2024xxc,
    author = "Chakraborty, Dibyendu and Chatterjee, Arindam and Kaushik, Ayushi and Nishiwaki, Kenji",
    title = "{Prospects of five-dimensional L{\ensuremath{\mu}}-L{\ensuremath{\tau}} gauge interactions in the light of elastic neutrino-electron scatterings: The scope of the DUNE near detector}",
    eprint = "2407.20615",
    archivePrefix = "arXiv",
    primaryClass = "hep-ph",
    doi = "10.1103/PhysRevD.110.095030",
    journal = "Phys. Rev. D",
    volume = "110",
    number = "9",
    pages = "095030",
    year = "2024"
}

@article{Chakraborty:2025jbd,
    author = "Chakraborty, Dibyendu and Chatterjee, Arindam and Kaushik, Ayushi and Nishiwaki, Kenji",
    title = "{Muon beam dump experiments probe five-dimensional nature of U(1)L{\ensuremath{\mu}}-L{\ensuremath{\tau}}}",
    eprint = "2510.25613",
    archivePrefix = "arXiv",
    primaryClass = "hep-ph",
    doi = "10.1103/1rmw-ms5d",
    journal = "Phys. Rev. D",
    volume = "113",
    number = "11",
    pages = "115038",
    year = "2026"
}

@article{Chakraborty:2026ocj,
    author = "Chakraborty, Dibyendu and Chatterjee, Arindam and Datta, AseshKrishna and Kaushik, Ayushi and Nishiwaki, Kenji",
    title = "{Aspects of a Five-Dimensional $U(1)_{L_\mu- L_\tau}$ Model at Future Muon-Based Colliders}",
    eprint = "2604.05041",
    archivePrefix = "arXiv",
    primaryClass = "hep-ph",
    month = "4",
    year = "2026"
}

@article{Schramm:1980xv,
    author = "Schramm, David N. and Steigman, Gary",
    title = "{Relic Neutrinos and the Density of the Universe}",
    reportNumber = "BA-80-19",
    doi = "10.1086/158559",
    journal = "Astrophys. J.",
    volume = "243",
    pages = "1",
    year = "1981"
}

@inproceedings{Naab:2023xcz,
    author = "Naab, Richard and Ganster, Erik and Zhang, Zelong",
    collaboration = "IceCube",
    title = "{Measurement of the astrophysical diffuse neutrino flux in a combined fit of IceCube's high energy neutrino data}",
    booktitle = "{38th International Cosmic Ray Conference}",
    eprint = "2308.00191",
    archivePrefix = "arXiv",
    primaryClass = "astro-ph.HE",
    reportNumber = "PoS-ICRC2023-1064",
    month = "7",
    year = "2023"
}

@article{Asai:2020qax,
    author = "Asai, Kento and Hamaguchi, Koichi and Nagata, Natsumi and Tseng, Shih-Yen",
    title = "{Leptogenesis in the minimal gauged U(1)$_{L_\mu-L_\tau}$  model and the sign of the cosmological baryon asymmetry}",
    eprint = "2005.01039",
    archivePrefix = "arXiv",
    primaryClass = "hep-ph",
    reportNumber = "IPMU-20-0048",
    doi = "10.1088/1475-7516/2020/11/013",
    journal = "JCAP",
    volume = "11",
    pages = "013",
    year = "2020"
}

@article{Ibe:2026yei,
    author = "Ibe, Masahiro and Miyamoto, Jun and Shirai, Satoshi",
    title = "{Radiative Breaking of Two-Zero Neutrino Mass Minors: Revisiting the $\mathrm{U}(1)_{L_\mu-L_\tau}$ Model}",
    eprint = "2607.06102",
    archivePrefix = "arXiv",
    primaryClass = "hep-ph",
    reportNumber = "IPMU26-0028",
    month = "7",
    year = "2026"
}

@article{delaVega:2024pbk,
    author = "de la Vega, Leon M. G. and Peinado, Eduardo and Wudka, Jose",
    title = "{L{\ensuremath{\mu}}-L{\ensuremath{\tau}} solution to the IceCube ultrahigh-energy neutrino deficit in light of NA64}",
    eprint = "2406.19968",
    archivePrefix = "arXiv",
    primaryClass = "hep-ph",
    doi = "10.1103/PhysRevD.110.095032",
    journal = "Phys. Rev. D",
    volume = "110",
    number = "9",
    pages = "095032",
    year = "2024"
}

@article{Araki:2014ona,
    author = "Araki, Takeshi and Kaneko, Fumihiro and Konishi, Yasufumi and Ota, Toshihiko and Sato, Joe and Shimomura, Takashi",
    title = "{Cosmic neutrino spectrum and the muon anomalous magnetic moment in the gauged $L_{\mu}-L_{\tau}$ model}",
    eprint = "1409.4180",
    archivePrefix = "arXiv",
    primaryClass = "hep-ph",
    reportNumber = "STUPP-14-219",
    doi = "10.1103/PhysRevD.91.037301",
    journal = "Phys. Rev. D",
    volume = "91",
    number = "3",
    pages = "037301",
    year = "2015"
}

@article{Chauhan:2018dkd,
    author = "Chauhan, Bhavesh and Mohanty, Subhendra",
    title = "{Signature of light sterile neutrinos at IceCube}",
    eprint = "1808.04774",
    archivePrefix = "arXiv",
    primaryClass = "hep-ph",
    doi = "10.1103/PhysRevD.98.083021",
    journal = "Phys. Rev. D",
    volume = "98",
    number = "8",
    pages = "083021",
    year = "2018"
}

@article{Barenboim:2019tux,
    author = "Barenboim, Gabriela and Denton, Peter B. and Oldengott, Isabel M.",
    title = "{Constraints on inflation with an extended neutrino sector}",
    eprint = "1903.02036",
    archivePrefix = "arXiv",
    primaryClass = "astro-ph.CO",
    doi = "10.1103/PhysRevD.99.083515",
    journal = "Phys. Rev. D",
    volume = "99",
    number = "8",
    pages = "083515",
    year = "2019"
}

@article{Bustamante:2020mep,
    author = "Bustamante, Mauricio and Rosenstr{\o}m, Charlotte and Shalgar, Shashank and Tamborra, Irene",
    title = "{Bounds on secret neutrino interactions from high-energy astrophysical neutrinos}",
    eprint = "2001.04994",
    archivePrefix = "arXiv",
    primaryClass = "astro-ph.HE",
    doi = "10.1103/PhysRevD.101.123024",
    journal = "Phys. Rev. D",
    volume = "101",
    number = "12",
    pages = "123024",
    year = "2020"
}

@article{Hyde:2023eph,
    author = "Hyde, Jeffrey M.",
    title = "{Constraints on Neutrino Self-Interactions from IceCube Observation of NGC 1068}",
    eprint = "2307.02361",
    archivePrefix = "arXiv",
    primaryClass = "hep-ph",
    month = "7",
    year = "2023"
}

@article{Araki:2015mya,
    author = "Araki, Takeshi and Kaneko, Fumihiro and Ota, Toshihiko and Sato, Joe and Shimomura, Takashi",
    title = "{MeV scale leptonic force for cosmic neutrino spectrum and muon anomalous magnetic moment}",
    eprint = "1508.07471",
    archivePrefix = "arXiv",
    primaryClass = "hep-ph",
    reportNumber = "UME-PP-002, STUPP-15-223",
    doi = "10.1103/PhysRevD.93.013014",
    journal = "Phys. Rev. D",
    volume = "93",
    number = "1",
    pages = "013014",
    year = "2016"
}

@article{Boccaletti:2024guq,
    author = "Boccaletti, A. and others",
    title = "{Hybrid calculation of hadronic vacuum polarization in muon g {\ensuremath{-}} 2 to 0.48{\%}}",
    eprint = "2407.10913",
    archivePrefix = "arXiv",
    primaryClass = "hep-lat",
    doi = "10.1038/s41586-026-10449-z",
    journal = "Nature",
    volume = "653",
    number = "8114",
    pages = "373--377",
    year = "2026"
}

\end{document}